\documentclass[12pt,preprint]{aastex}

\usepackage{a4}

\usepackage{graphicx}

\newcommand{\super}[1]{\ensuremath{^\textrm{\scriptsize{#1}}}}
\newcommand{\sub}[1]{\ensuremath{_\textrm{\scriptsize{#1}}}}

\newcommand{\vgsr}{$V$\sub{GSR}}
\newcommand{\meanvgsr}{$\langle$$V$\sub{GSR}$\rangle$}
\newcommand{\kms}{km s$^{-1}$}
\newcommand{\degree}{\super{o}}

\newcommand{\meanfeh}{$\langle$[Fe/H]$\rangle$}

\slugcomment{\emph{Accepted for Publication in ApJ}}

\shorttitle{Extending the VSS with SEKBO}
\shortauthors{Prior et al.}

\begin{document}

\title{Extending the Virgo Stellar Stream\\
    with SEKBO Survey RR Lyrae Stars}

\author{Sayuri L. Prior, G. S. Da Costa, Stefan C. Keller and Simon J. Murphy}
\affil{Research School of Astronomy and Astrophysics, Australian National University, Cotter Road, Weston Creek, Canberra, ACT 2611, Australia}

\begin{abstract}
A subset of the RR Lyrae (RRL) candidates identified from the Southern
Edgeworth-Kuiper Belt Object (SEKBO) survey data has been followed up
photometrically ($n=106$) and spectroscopically ($n=51$).  Period and light 
curve fitting reveals a $24 \pm 7 \%$ contamination of SEKBO survey data by 
non-RRLs.  This paper focuses on the region of the Virgo Stellar Stream (VSS), 
particularly on its extension to the south of the declination limits of the 
SDSS and of the QUEST RRL survey.  The distribution of radial velocities in 
the Galactic standard of rest frame (\vgsr) for the 11 RRLs observed in the 
VSS region has two apparent peaks.  The larger peak coincides with the four 
RRLs having \meanvgsr = $127 \pm 10$ \kms{} and dispersion $\sigma = 27$ \kms, 
marginally larger than that expected from the errors alone.  The two type $ab$
RRLs in this group have \meanfeh{} = $-1.95 \pm 0.1$.  Both the radial
velocities and metal abundances are consistent with membership in the
VSS\@. The second velocity peak, which occurs at \meanvgsr = $-175 \pm 10$
\kms{} may indicate the presence of stars from the Sgr leading tidal tail, 
which is expected to have large negative velocities in this region.  
We explore the spatial extent of the VSS by constructing luminosity
functions from the SEKBO data and comparing them to data synthesized with 
the Besan\c{c}on Galactic model.  Analysis of the excess over the model 
predictions reveals the VSS as a large ($\sim$760 deg$^2$) overdensity 
centered at roughly (RA, Dec) $\sim$ (186\degree, $-4$\degree), spanning a 
length of $\sim$15 kpc in projection, assuming a heliocentric distance of 
19 kpc.  The data reveal for the first time the more southern regions of the
stream and trace it to Dec $\approx -15$\degree\ and Galactic latitudes
as low as $b \approx 45$\degree.
\end{abstract}

% ApJ style

\section{Introduction}

It is now widely accepted that galaxies are at least partly formed by
a prolonged, chaotic aggregation of independent, protogalactic
fragments, consistent with the proposal of \citet{SZ78}.  This
conceptualization is in line with the currently favored $\Lambda$ cold
dark matter ($\Lambda$CDM) cosmologies which postulate that galaxy
formation is a consequence of the hierarchical assembly of subgalactic
dark halos and the subsequent accretion of cooled baryonic gas (see,
for example, \citealt{CB01} and references therein).  The hierarchical
picture stands in contrast to the formerly held conceptualization that
a rapid, free-fall collapse of an isolated pre-galactic cloud was the
crucial galaxy-forming event \citep{ELS62}.  Nevertheless, the
observed dual properties of the Galactic halo (e.g.\ in terms of
density, kinematics and age) now point to a possible combined scenario
wherein the inner and outer halo regions were formed by different
mechanisms (see, for example, \citealt{CBL07}).

Of all the Galactic components, the outer halo presents arguably the
best opportunity for probing its formation due to its remoteness and
relative quiescence.  In order to gauge quantitatively the relative
importance of the accretion mechanism in halo formation, \citet{BZB08}
compare the level of substructure present in Sloan Digital Sky Survey
(SDSS) data to simulations, and find that the data are consistent with
a halo constructed entirely from disrupted satellite remnants.  Direct
evidence of systems currently undergoing disruption have indeed been
found, with the prime example being the Sagittarius (Sgr) dwarf galaxy,
which is located a mere 16 kpc from the Galactic center.  From the
time of its discovery, the elongated morphology of Sgr, pointing towards the
Galactic center, has been taken as evidence for strong, ongoing tidal
disruption \citep{IGI94}.  A combination of observations have
subsequently found the debris from the interaction to wrap around the
sky \citep{MSW03,NYG03,BZE06}, making it the most significant known
contributor to the Galactic halo.

A number of other streams and groups have been identified in the halo.
Examples include the Monoceros Stream \citep{NYR02,YNG03} which
surrounds the Galaxy in a giant ring \citep{IIL03}, and the
Hercules-Aquila Cloud \citep{BEB07} which extends above and below the
Galactic plane and stretches $\sim$80\degree{} in longitude.  Another
significant feature was discovered in Quasar Equatorial Survey Team
(QUEST) data as an overdensity of RR Lyrae stars
\citep{VZA01,VZA04,VZ06}, and in SDSS data as an excess of F-type main
sequence stars \citep{NYR02,NYC07}, in the  direction of the Virgo
constellation.  QUEST, dubbing the feature the ``12\super{h}.4
clump'', estimated its heliocentric distance as  $\sim$19 kpc, centered
at RA $\sim 186$\degree{}.  It was found to span RA $\sim
175$\degree--$200$\degree{} and the Dec range of the QUEST survey
($-2.3$\degree--$0.0$\degree).  Subsequent radial velocity
measurements by \citet{DZV06} of a subset of the clump revealed a
common velocity in the Galactic standard of rest frame (\vgsr) of
$100$ \kms{} and a dispersion of $\sigma = 17$ \kms, slightly smaller
than the average error of the measurements.  Using SDSS data, they
estimated the feature to cover at least $106$ deg\super{2} of sky and
suggested the name, ``Virgo Stellar Stream'' (VSS).

Using photometric parallaxes of SDSS stars, \citet{JIB08} identified
the ``Virgo Overdensity'' (VOD) as a large ($\sim$1000 deg\super{2}),
diffuse overdensity in the same direction as the VSS, but at distances
$\sim$6--20 kpc.  In a recent paper, \citet{VJZ08} provided additional
information in this region of the sky for distances less than 13 kpc.
They found that the VSS extends to distances as short as 12 kpc, in
comparison to its previous detection at 19 kpc.  In spite of the
differences in reported distances and velocities, \citet{NYC07}
suggested that all the observed overdensities in Virgo may be the same
feature.  The possibility exists that the terms may not be
interchangeable, but for simplicity, the current paper hereafter
refers to the feature as the VSS.  The association of the VSS with Sgr
debris was hypothesized by \citet{MPJ07} who showed that
\citeauthor{LJM05}'s \citeyearpar{LJM05} model of the Sgr leading
tidal tail passes through the region of the VSS\@.  However, the model
predicts highly negative radial velocities for Sgr stars in this
region, contrary to the observations of \citet{DZV06} (see above) and
\citet{NYC07} who find the most significant peak at \vgsr = $130 \pm
10$ \kms.  The model also predicts a relatively low density of Sgr
debris in this region which is at odds with the significance of the
observed overdensity. \citeauthor{NYC07} also note that the VSS is not
spatially coincident with the main part of the Sgr leading tidal tail,
but that the features do significantly overlap.

The foregoing discussion is only a brief summary of the findings
relating to the overdensity in Virgo, but it highlights the
considerable uncertainty that remains regarding its spatial form and
origin.  From the results of \citet{BZE06} and \citet{NYC07}, it is
highly probable that the center of the VSS in fact lies to the south
of the regions mapped by SDSS and QUEST\@.  This region is covered by
the Southern Edgeworth-Kuiper Belt Object (SEKBO, \citealt{MSA03})
survey, which is discussed in more detail in \S\ref{targetselection}.
\citet{KMP08} produced a list of over 2000 RR Lyrae (RRL) variable
star candidates from this data set and analyzed their spatial
distribution.  Among other overdensities, they identify two clumps in
the region of the VSS\@.  Clump 1, at a heliocentric distance of 16 kpc,
is located $\sim$8\degree{} south-east of the VSS centre identified by
\citet{DZV06}, while Clump 2 is at a distance of 19 kpc and located
$\sim$16\degree{} further to the south-east.  The current study
follows up a subset of these RRLs.  After confirming their RRL
classification with photometric observations (\S\ref{obs}), radial
velocities from spectra enable us to determine whether they belong to
the VSS (\S\ref{rv}).  Metal abundances are also calculated
(\S\ref{metal}).  To obtain further information regarding the spatial
extent of the stream, a wider stellar population from the SEKBO survey
data set is examined for signs of an excess of stars in the region of
interest (\S\ref{lf}).  The targets selected for follow-up also
include clumps of RRL candidates in regions overlapping Sgr debris,
though the results for these stars are deferred to a subsequent paper.
In addition, a few smaller apparent spatial groupings were also
targeted for spectroscopic follow-up to investigate whether they are
associated with other substructures in the halo.

\section{Observations and Data Reduction}
\label{obs}

\subsection{Target Selection} 
\label{targetselection}

Targets were selected from a list of 2016 candidates produced by
\citet{KMP08} who searched the Southern Edgeworth-Kuiper Belt Object
survey data for RRLs.  The SEKBO survey was conducted on the $50''$
telescope at Mount Stromlo Observatory between January 2000 and 2003
and covered a 10\degree{} wide band following the ecliptic (1675
deg$^2$ of imaging data).  Two filters (`blue': 455--590 nm and `red':
615--775 nm) were used simultaneously, with typically a set of three
300 s observations obtained, separated by $\sim$4 hrs and $\sim$1--7
days.  In order to select candidates, a score was constructed that
measured how well an object matched the expected properties of an RRL
(i.e.\ in terms of its color and variability).  Analysis showed that
this procedure produced a candidate list with completeness for RR$ab$
$\sim$60\% for $V<18.5$, falling to 25\% by $V=19.5$.  Further details
can be found in \citet{KMP08} (hereafter KMP08).

\begin{figure}[htb]     
       \centering
       \epsscale{0.6}
        % Online version in color :
        %\plotone{f1_online.eps}
        %%BoundingBox: 162 280 430 552
        \plotone{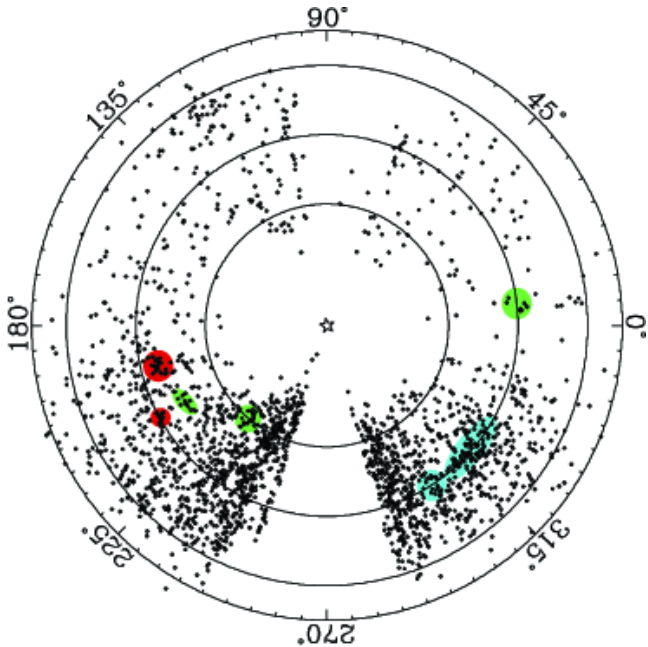}
        \caption[The heliocentric radial distribution of RRL
        candidates from the SEKBO survey in ecliptic longitude (from
        KMP08).]{The radial distribution of RRL candidates from the
        SEKBO survey in ecliptic longitude, $\lambda$ (from KMP08).
        The concentric circles are at $V_0$=15, 17 and 19.  Shaded areas
        represent regions of apparent overdensity which were followed
        up in the current study.  Red ($\lambda \sim 195$\degree{} and
        $210$\degree): VSS region; Blue ($\lambda \sim 305$\degree{}
        and $315$\degree): Sgr region; Green ($\lambda \sim
        210$\degree, $230$\degree{} and $5$\degree): 14h, 16 h, and 0
        h regions.}
      
        \label{radial}  
\end{figure}

The heliocentric radial distribution of RRL candidates from the SEKBO
survey is displayed in Figure \ref{radial}.  From this set of
candidates, several apparent clumps of stars were targeted for
follow-up.  Firstly, as the SEKBO survey region overlaps that of the
QUEST survey in the vicinity of the VSS, the possibility existed to
not only recover previously identified VSS members, but also to gain
further information regarding the spatial extent of the stream.  A
selection of 8 of the 13 RRLs from the candidate list falling within
the RA range $183$\degree--$192$\degree{} and $V_0$ range
$16.6$--$17.2$ was targeted for observation (`VOD Clump 1' in KMP08).
A second clump located at RA $\sim 206$\degree{} at a similar distance
was also targeted (`VOD Clump 2' in KMP08).  In addition, apparent
spatial groupings of stars at (RA, $V$) of  $\sim$  (14 h, 16 mag),
(16 h, 15 mag), (20 h, 17 mag), (21.5 h, 17 mag) and (0 h, 17 mag)
were targeted.  Note that the spatial position of the 20 and 21.5 h
stars overlaps the expected position of the trailing arm of the Sgr
debris stream.

%-------------------- radial --------------------------%
%------------------------------------------------------%

% contamination targets
In addition to the 51 RRLs from the clumps which were targeted for
spectroscopic follow-up, a selection of a further 55 candidates over a
wide range of RAs was targeted for photometric follow-up in order to
elucidate the nature of the contamination of the SEKBO RRL candidate
sample by non-RRLs.  These targets varied in magnitude between $V$ =
15 and 19.5.  This set of targets included a selection of `red'
variable objects to investigate whether the adopted dereddened color
cutoff of $(V-R)_0 = 0.3$ for RRL candidates was appropriate.  The
number of spectroscopy and photometry targets in each region is
summarized in Table \ref{targets}.

\subsection{Photometry}

\begin{figure}[htb]     
        \centering
        \epsscale{0.5}
        \plotone{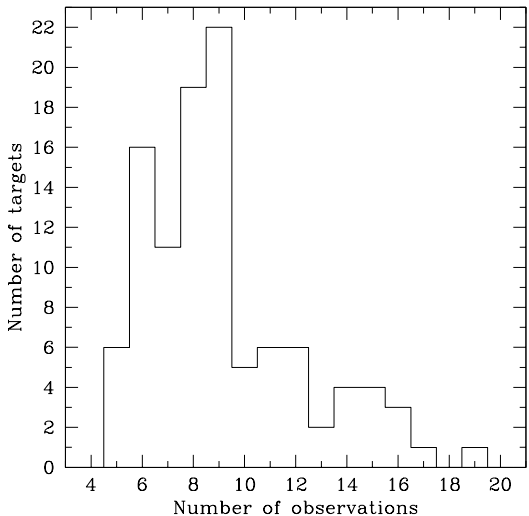}
        \caption[The distribution of number of observations for imaged
        candidates.]{The distribution of number of observations for
        imaged candidates.}  
        \label{nobs}    
\end{figure}

% observations
Observations were made with the Australian National University (ANU)
$40''$ telescope at Siding Spring Observatory (SSO) over six six-night
runs between November 2006 and October 2007.  The target was centered
on one of the Wide Field Imager's eight CCDs.  The $V$ filter was
used, with exposure times ranging from 120 to 600 s depending on the
target magnitude.  The total number of observations for each target
over the observing runs ranged from 5 to 19, with an average of 9
observations per target (see Figure \ref{nobs}).

%--------------------- nobs ----------------------%
%-------------------------------------------------%

\begin{figure}[htb]
        \centering 
        \epsscale{1}
        \plottwo{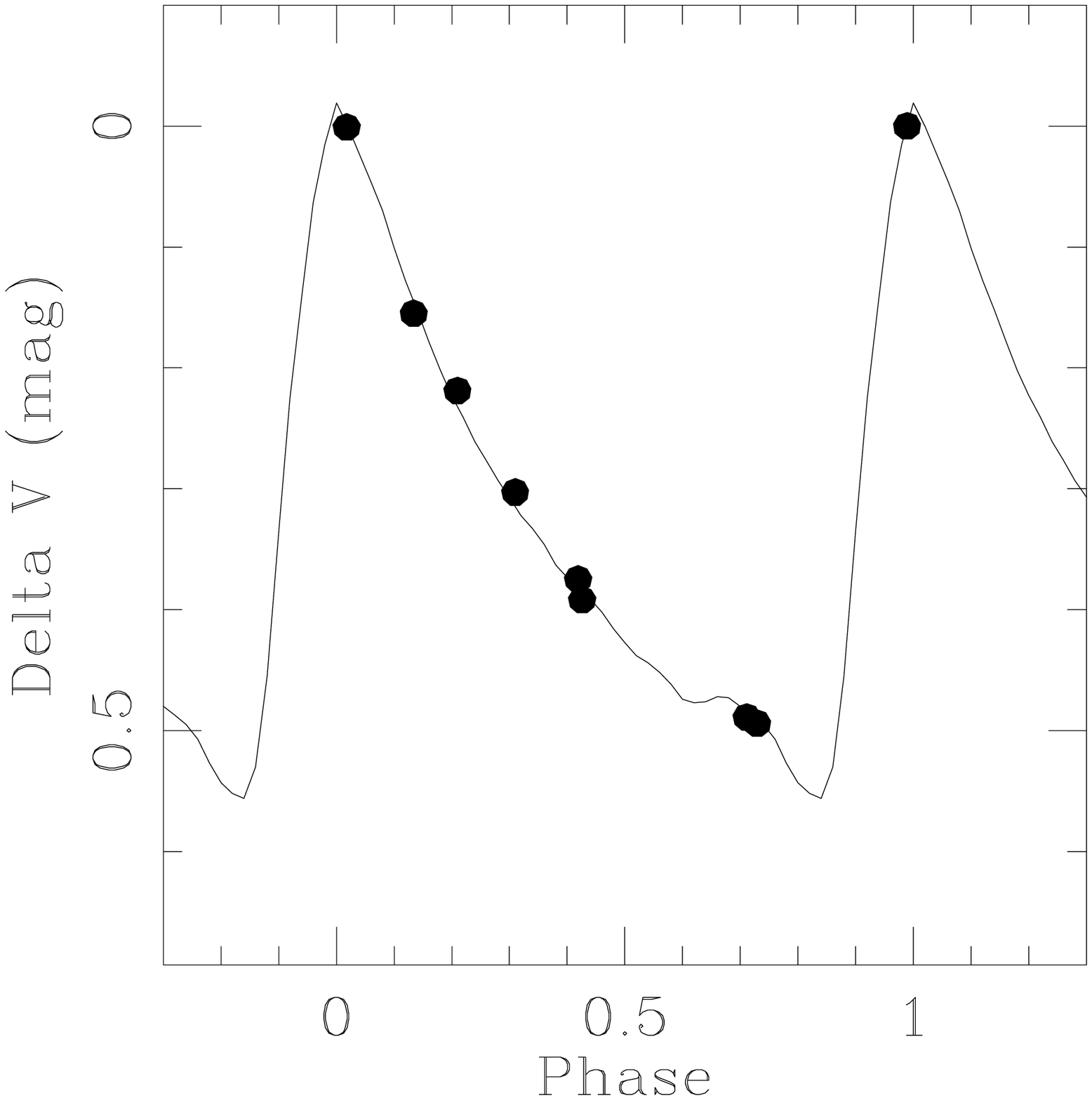}{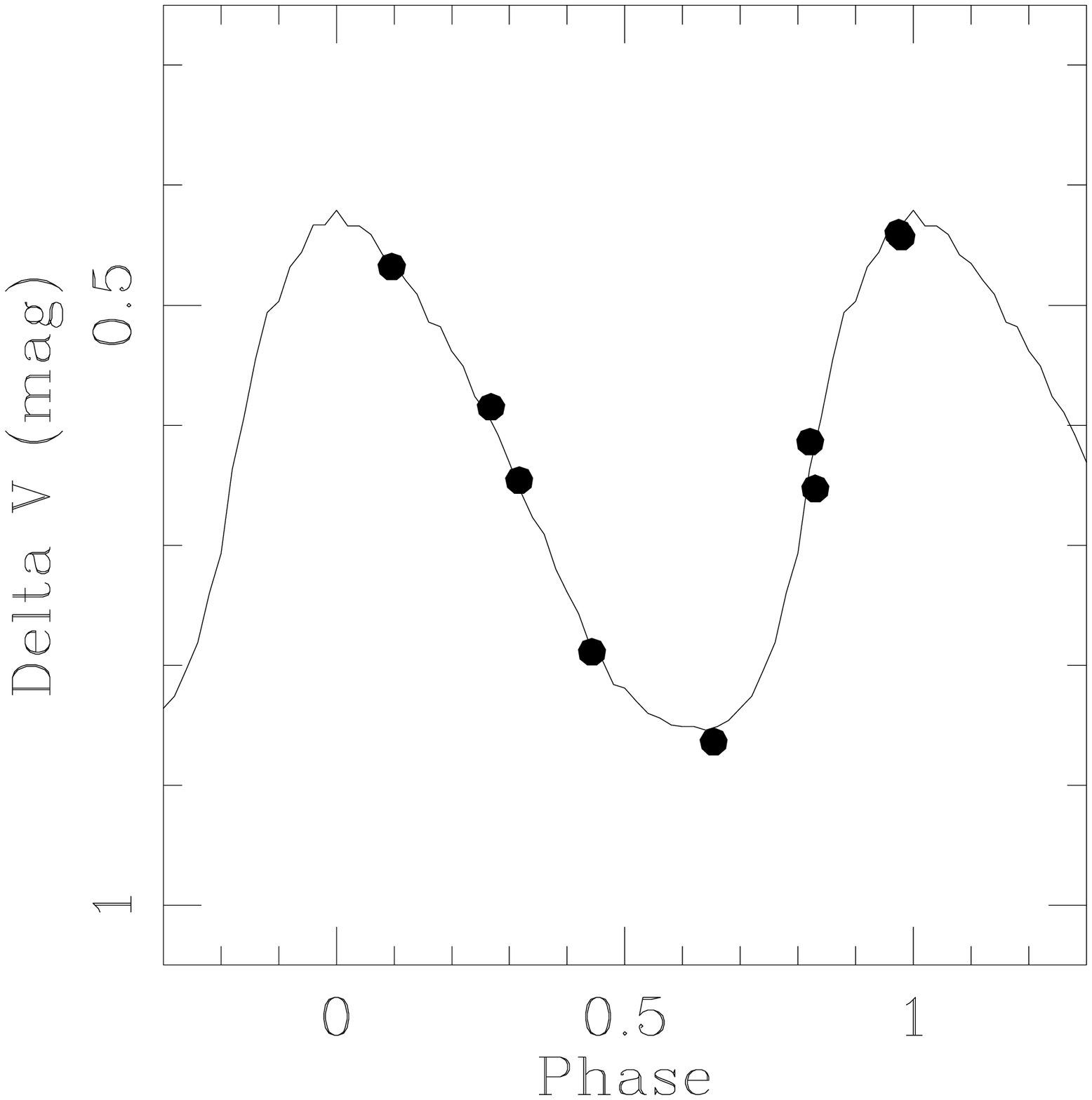}
        \caption[Phased $V$ mag differences between a candidate and
        nearby comparison stars.]{Phased $V$ mag differences between a
        candidate and nearby comparison stars.  From the tight fit of
        the data points to the template light curves \textit{(solid
        line)}, we can be very confident that these two candidates are
        indeed RR Lyrae stars. \textit{Left: } RR$ab$ with period =
        0.583 d.  \textit{Right: } RR$c$ with period = 0.358 d.} 
        \label{lightcurve}
\end{figure}

% data reduction
The data were overscan subtracted, trimmed, bias subtracted and
flatfielded with twilight sky flats using standard IRAF procedures.
Aperture photometry was then performed on each target as well as on an
ensemble of nearby comparison stars, yielding an average differential
magnitude for each target at each epoch.   This series of differential
magnitudes for the target was subsequently entered into Andrew
Layden's period-fitting routine (\citealt{LS00} and references
therein) along with the mid-exposure heliocentric Julian dates of the
observations.  Layden's routine identifies the most likely period by
fitting the photometry of the variable star with 10 templates
(including 7 RRL templates) and performing a $\chi^{2}$ minimization.
See \citet{PAJ02} for a more detailed description of the method.  The
best obtained period was then entered into Layden's light
curve-fitting routine.  Example candidate light curves are displayed
in Figure \ref{lightcurve}.    

%------------------- lightcurve --------------------%
%---------------------------------------------------%

\begin{figure}[htb]
        \centering 
        \epsscale{0.7}
        \plotone{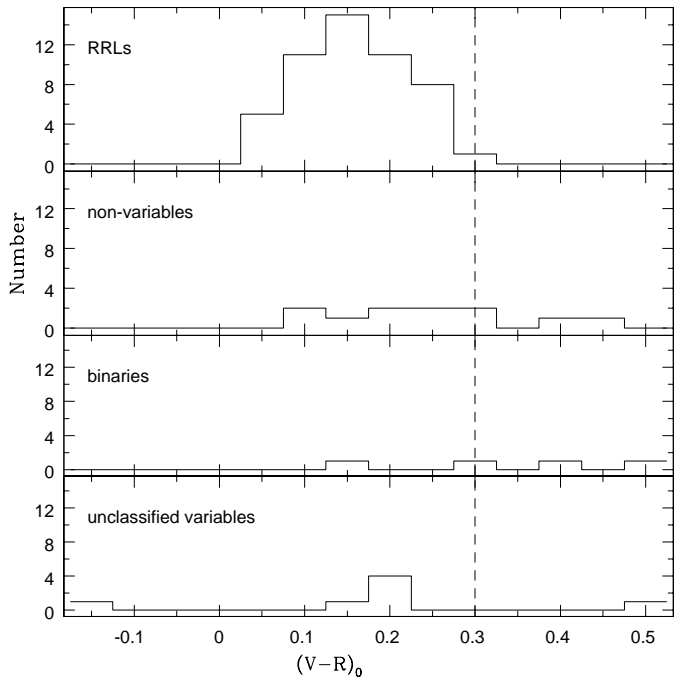}
        \caption[The classification of RRL candidates based on
        follow-up photometry.]{The classification of RRL candidates
        based on follow-up photometry.  The dashed line indicates the
        adopted color cutoff of $(V-R)_0<0.3$ for RRL candidates in
        the SEKBO survey.}   
        \label{contam}
\end{figure}

One important application of the analysis of this photometry was to
determine which candidates were indeed RRL stars and which were
spurious detections.  Of the 73 stars observed at least 8 times, 11
had magnitudes which did not vary significantly over the observations,
4 were classified as binary stars and 7 showed variability but did not
appear to have periods and light curves corresponding to either RRLs
or binary stars.  Figure \ref{contam} shows the number of candidates
falling into each classification category as a function of $(V-R)_0$.

Given that there are 51 RRLs and 16 non-RRLs within the color
selection, this indicates that the procedure KMP08 used to identify
RRLs from the SEKBO survey data has a contamination rate of 24 $\pm$
7\%, where the uncertainty has been calculated using Poisson
statistics.  Table \ref{photdata} summarizes the photometric data,
including the classification, period and fitted $V$ amplitudes (where
applicable) for all targets.

%-------------------- contam ----------------------%
%--------------------------------------------------%

\subsection{Spectroscopy}

% observations
Observations were made with the ANU 2.3 m telescope at SSO in runs
that were concurrent with those on the $40''$ telescope.  The blue arm
of the Double Beam Spectrograph was used with a $2''$ slit and the
600B grating, giving a resolution of $\sim$2 \AA.  The spectra were
centered on 4350 \AA{} so that the Ca II K line and Hydrogen lines such
as H$\delta$, H$\gamma$ and H$\beta$ could be observed.  Exposure
times were chosen to give a signal-to-noise of $\sim$20 and varied
between 900 and 3000 s.  Each target was observed between one and four
times, spread throughout a given observing run.  Each observation was
accompanied by a comparison Cu Ar lamp exposure.  Radial velocity
standards of similar spectral type were also observed for use as
cross-correlation templates (see \S \ref{rv}) as well as standards for
\citeauthor{Layden94}'s \citeyearpar{Layden94} pseudo-equivalent width
system (see \S \ref{metal}).  A small number of bright RRLs from
\citet{Layden94} were also observed to serve as additional
cross-correlation templates.   

\begin{figure}[htb]
        \centering 
        \epsscale{0.7}
        \plotone{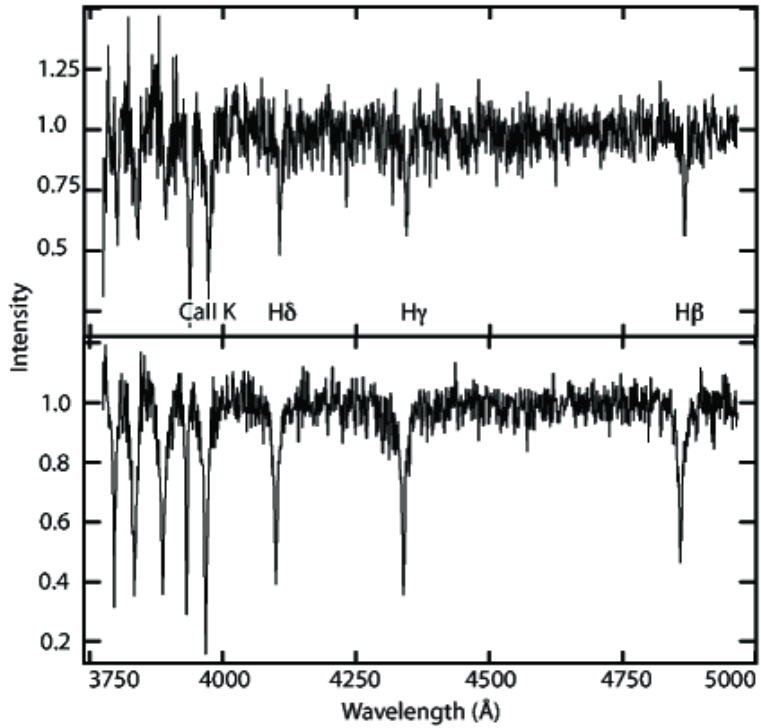}
        \caption{Examples of poor quality \textit{(top)} and good
        quality \textit{(bottom)} normalized spectra.} 
        \label{spectra}
\end{figure}

% data reduction
The data were overscan subtracted and trimmed using standard IRAF
procedures.  Bias subtraction and flatfielding were not performed as
they only served to add noise to the data.  Wavelength calibration was
performed over the range 3500--4965 \AA{} using 4th--8th order
legendre polynomials and 26--28 spectral lines, yielding dispersion
solutions with RMS $\sim$ 0.05 \AA.  Example spectra (after continuum
fitting as described in \S \ref{metal}) are displayed in Figure
\ref{spectra} and demonstrate the range in quality obtained.

%%BoundingBox: 125 411 434 707
%-------------------- spectra ----------------------%
%--------------------------------------------------%
  
%\input{rv}

\section{Radial Velocities}
\label{rv}

% fxcor
Radial velocities were determined using the IRAF task \textsc{fxcor}
which performs Fourier cross-correlations between spectra of the
target star and chosen template stars.  For each observing run, we
chose 5--6 stars to use as templates from amongst the radial velocity
standards and the bright RRLs from \citet{Layden94}.  The selection
process was guided by the fact that relative velocities between
template and target stars were most precise when the two stars were of
similar spectral type.  Subsequently, heliocentric corrections were
made to remove the component of the observed velocity due to the
Earth's motion around the Sun.    

\begin{figure}[htbp]    
        \centering
        \epsscale{0.8}
        \plotone{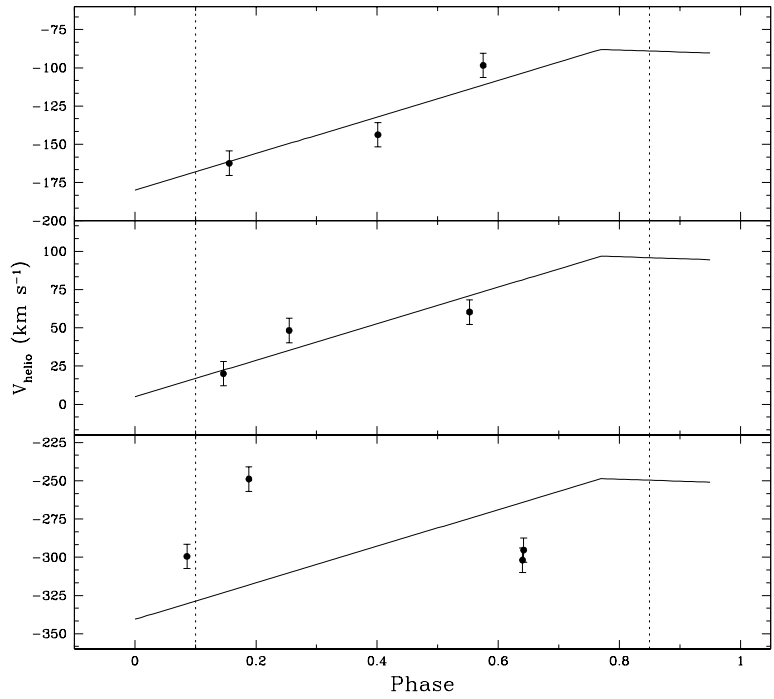}
        \caption[Fits of the radial velocity template of X Ari to
        three type $ab$ RRLs in our sample.]{Fits of the radial
        velocity template of X Ari \textit{(solid line)} to three type
        $ab$ RRLs in our sample, each observed at three phases.  The
        upper two plots are examples of excellent fits, while the
        lowest plot is an example of a poor fit.  Note that data
        points with phase $<$ 0.1 or phase $>$ 0.85 are excluded in
        the fit.  The systemic velocity corresponds to phase 0.5.}  
        \label{vphase}  
\end{figure}

% RRab: X Ari
Observed radial velocities of an RR$ab$ star can vary up to $\pm$50
\kms{} from its systemic velocity as a function of phase.  In order to
correct the observed velocities to systemic velocities we first
determined the phase of each observation using the ephemerides based
on the best period obtained from our photometric data.
\citeauthor{Layden94}'s \citeyearpar{Layden94} parametrization of the
velocity curve for the RRL star X Ari (measured by \citet{Oke66} from
the H$\gamma$ line) was then used to determine the systemic
velocity. Figure \ref{vphase} shows example fits of the curve to our
observed data.  Based on X Ari, the systemic velocity is taken to
occur at phase 0.5.  Note that because the form of the discontinuity
in the light curve near maximum light varies amongst RRLs, only phases
between 0.1 and 0.85 were used in the fit.  The average RMS of the
fits was 18 \kms{}, which we use as an estimate of the uncertainty in
the conversion from observed velocities to systemic velocities.  We
then combine this with the average uncertainty in the radial velocity
zeropoints across the observing runs ($\pm 7$ \kms) and the
uncertainty in the cross-correlations ($\pm 4$ \kms).  The latter was
quantified by calculating, for each observation, the standard
deviation of the velocities obtained using the different templates,
then averaging over all observations.  The combination of these errors
then yields an overall uncertainty in the systemic radial velocities
of $\pm 20$ \kms.  For the one RRL in common with \citet{DZV06}, the
velocities agree within the combined errors.

%-------------------- vphase --------------------------%
%------------------------------------------------------%

% RRc
Of the 51 spectroscopic targets, 13 were type $c$ RRLs. For these
stars, velocity data for DH Peg and T Sex \citep{LJ89,JCL88} were used
to create a template velocity curve.  However, given that the template
yielded uncertainties in systemic velocities of up to 30 \kms{} when
fitted to our data, we opted to instead use the average radial
velocity as our estimate of the systemic velocity.  Taking into
consideration the precision of our measurements and the fact that
observed radial velocities of an RR$c$ star at different phases only
vary by up to $\pm$30 \kms{} from its systemic velocity, one would not
expect a fit to a template curve to provide an improved constraint on
the systemic velocity.  This was indeed evidenced by the large scatter
in the residuals of the template fit to the data as a function of
phase.  The average radial velocity was also used as an estimate of
the systemic velocity for the four type $ab$ RRLs having all
observations at phases less than 0.1 or greater than 0.85.  This
non-optimal method likely produced inflated radial velocity
uncertainties for these stars. 

% Vgsr
When investigating Galactic substructures, it is useful to consider
radial velocities in a frame of reference which is at rest with
respect to the Galactic center. The heliocentric radial velocities
($V$\sub{helio}) calculated as described above were thus transformed
to Galactic standard of rest frame velocities (\vgsr{}), thereby
removing the effects of the Sun's peculiar motion (assumed to be
$(u,v,w) = (-9,11,6)$ \kms{} with respect to the local standard of
rest which has a rotation of 220 \kms{}).  The determined radial
velocities, $V$\sub{helio} and \vgsr, for RRLs in the VSS region and
the 14 h, 16 h and 0 h regions are given in Table \ref{spectro}.

\subsection{Virgo Stellar Stream region}

The spatial distribution and \vgsr{} of observed RRLs in VSS Clumps 1
and 2 are displayed in Figure \ref{12hrtno}.  Distances are based on
the assumption of $M_V$ = 0.56 and have an uncertainty of $\sim$7\%,
as described in KMP08.  This corresponds to an uncertainty of
approximately $\pm$1 kpc at a distance of 20 kpc.  RRLs observed
spectroscopically by \citet{DZV06} (hereafter DZV06) and classified as
VSS members are also included on the Figure to show the region where
the VSS was detected.  The four magenta points have $40$ \kms{} $<$
\vgsr{} $< 160$ \kms, the range within which DZV06 classified RRLs as
members of the VSS.  One star has previously been associated with the
VSS by DZV06 while the remaining three proposed members are new
discoveries.  They would suggest that the stream spans a much larger
declination range than previously estimated.  The four members have
\meanvgsr{} = $127 \pm 10$ \kms{} and dispersion $\sigma = 27$ \kms{{}
  which is only slightly larger than the measurement error of $\pm20$
  \kms.  Our value of \meanvgsr{} is somewhat higher than DZV06's
  value of $100\pm8$ \kms\ and is in better agreement with
  \citeauthor{NYC07}'s \citeyearpar{NYC07} value of $130\pm10$ \kms.

%-------------------- 12hrtno --------------------------%
\begin{figure}[htbp]    
        \centering
        \epsscale{0.8}
        \plotone{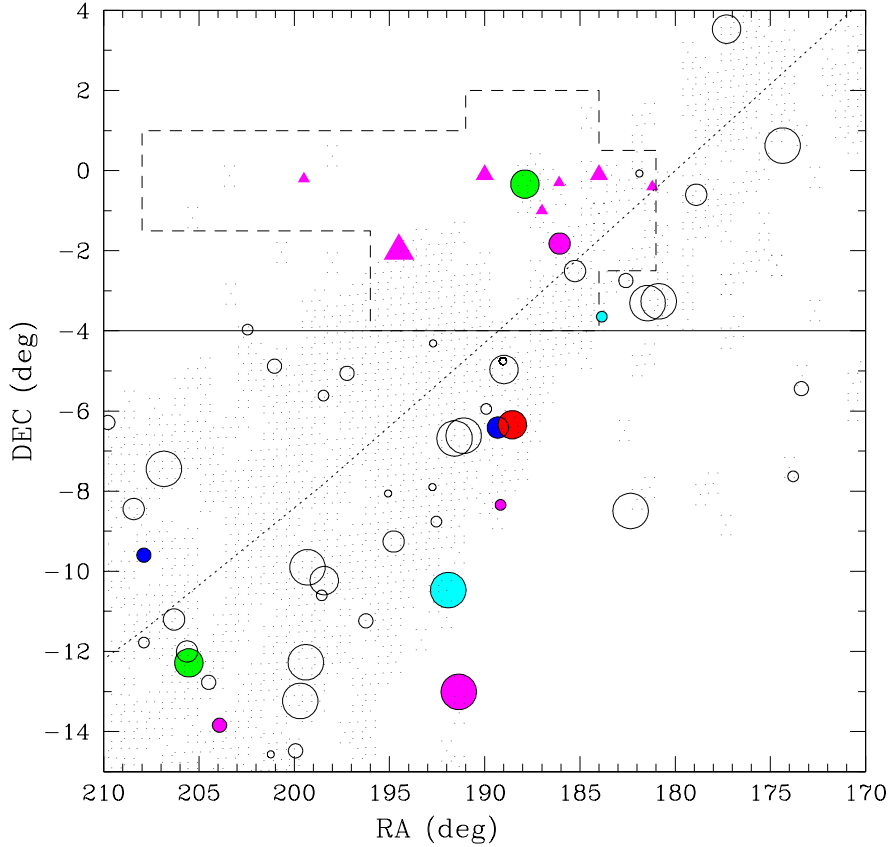}
        \caption[The spatial and \vgsr{} distribution of RRL
        candidates from the SEKBO survey and QUEST RRLs found by
        \citet{DZV06} to be VSS members.]{The spatial distribution of
        RRL candidates from the SEKBO survey \textit{(circles)} and
        QUEST RRLs found to be VSS members \textit{(triangles)}.
        Symbol size represents distance, ranging from 16 kpc
        \textit{(largest)} to 22 kpc \textit{(smallest)}.  Filled
        circles were observed spectroscopically and are color-coded
        according to \vgsr.  \textit{Blue:} less than $-160$ \kms{};
        \textit{Cyan:} $-160$ to $-40$ \kms{}; \textit{Green:} $-40$
        to 40 \kms{}; \textit{Magenta:} 40 to 160 \kms{} and
        \textit{Red:} greater than 160 \kms.  The dotted line is the
        ecliptic and the small dots represent the coverage of the
        SEKBO survey fields in this part of the sky.  The dashed box
        is the region where \citet{DZV06} identified the VSS as an
        excess of main-sequence stars of appropriate apparent
        magnitude.  The approximate southern limit of SDSS data in
        this region is shown by the solid line at Dec = $-4$\degree.}
        \label{12hrtno} 
\end{figure}
%------------------------------------------------------%

Figure \ref{genhistovss} is a generalized histogram of \vgsr{} in
which each observed value is represented by a normal distribution with
mean equal to the observed value and standard deviation comparable to
the uncertainty in the measurement.  Summing probability density
distributions over all observations then yields an estimate of the
true distribution of \vgsr{} which, unlike in standard histograms,
does not vary according to binning choice.  A random selection of halo
stars is expected to have a normal distribution of radial velocities
(e.g. \citealt{HMO01}) with a mean of $\sim$0 \kms{} and a standard
deviation of $\sim$100 \kms (e.g.\ \citealt{SGK04,BGK05}) (dotted line
on Figure \ref{genhistovss}).  Our data does not fit the expected
distribution well, though the difference did not reach statistical
significance in a Kolmogorov-Smirnov test due to the small sample
size.   

On visual inspection, however, there is a suggestion of a peak at
$\sim$130 \kms{} (VSS members) and a second at large negative velocities.  
The three stars contributing to this latter peak, for which 
\meanvgsr{} = $-175 \pm 10$ \kms{},  are similar in velocity (and general
spatial location) to the excess of stars
identified by \citet{NYC07} at \vgsr{} $= -168 \pm 10$ \kms{}
and (RA, Dec) $\approx$ (191\degree, $-8$\degree{}).  \citet{NYC07} did not suggest an
association for this peak, but it has the appropriate \vgsr{} to be
associated with the Sgr leading tidal tail, which is expected to have a highly
negative radial velocity at this RA (e.g.\ see modeling by
\citealt{LJM05}).  However, debate currently surrounds the question of
whether Sgr debris, coming from the north Galactic pole to the solar
neighborhood, is in fact densely located in this region
\citep{MPJ07,NYC07}. \citeauthor{NYC07} do concede, however, that the
VSS and Sgr streams overlap, so while it is not expected
to be dense enough in this region to account for the entire
overdensity in Virgo, it seems plausible that a portion could be
attributed to Sgr leading debris.  

%--------------------------- genhistovss --------------------------%
% epsffit -r 22 16 596 784 genhisto_bw.eps | epsffit -r 22 16 596 784 |
% epsffit -r 22 16 596 784 > genhisto_rot_bw.eps
% BB 22 44 500 320
\begin{figure}[htbp]    
        \centering
        \epsscale{1}
        \plotone{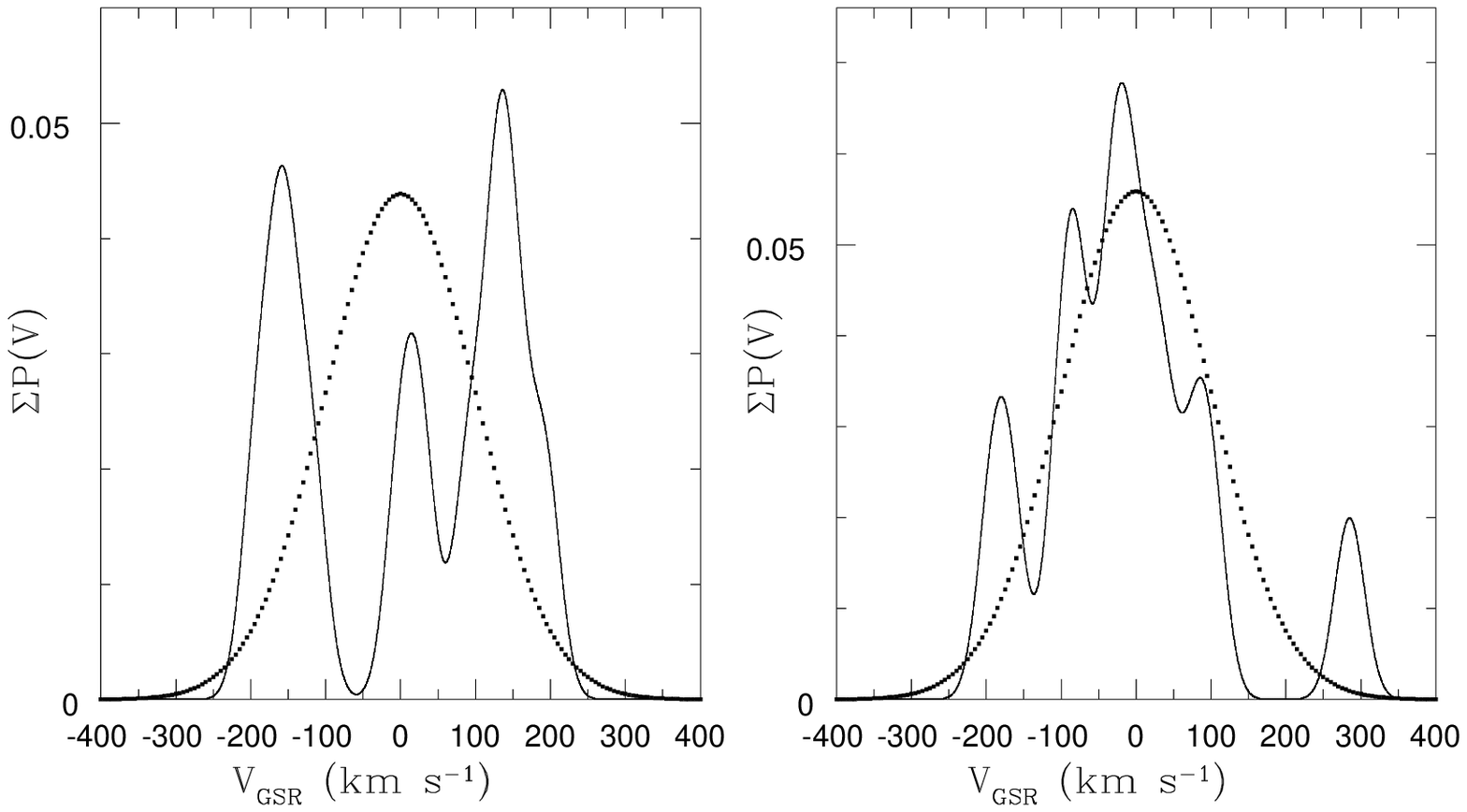}
        \caption[Generalized histogram of \vgsr{} for the eleven
        observed RRLs in the VSS region.]{Generalized histogram of
        \vgsr{} (with kernel of 20 \kms{}) for the eleven observed
        RRLs in the VSS region \textit{(left)} and for the fourteen
        RRLs in the 14 h, 16 h, and 0 h regions \textit{(right)}.
        Overplotted \textit{(dotted line)} is the expected
        distribution of \vgsr{} given a halo population in which
        velocities are normally distributed with \meanvgsr{} = 0
        \kms{} and $\sigma$ = 100 \kms{}.  The VSS can be seen in the
        left panel as the excess of stars at \vgsr{} $\sim 130$ \kms,
        while the excess at \vgsr{} $\sim -170$ \kms{} could
        potentially be related to leading Sgr debris.} 
        \label{genhistovss}     
\end{figure}
%------------------------------------------------------%

\subsection{Regions at RA 14 h, 16 h, and 0 h}

%% other clumps
No groupings of velocities were noted in the apparent spatial clumps
at 14 h, 16 h, and 0 h, though it should be noted that the small sample
sizes might have made any moving groups difficult to detect.  We note,
however, that the \vgsr{} distribution of RRLs in these regions (see
the right panel of Figure \ref{genhistovss}) more closely resembles
the expected normal distribution of halo stars than does the
distribution of RRLs in the VSS region.  As mentioned above, the
20 and 21.5~h regions overlap with the Sgr trailing debris stream and the 
results will be discussed in a separate paper.  
%The remainder of this paper
%focuses on the VSS region, particularly its extension to the
%south of the declination limits of the SDSS and the QUEST survey. 

%\input{metal}

\section{Metal Abundances}
\label{metal}

Metallicities ([Fe/H]) were calculated using the \citet{FR75} method
which is an analogue of \citeauthor{Preston59}'s
\citeyearpar{Preston59} classic $\Delta$S technique.  In the
\citeauthor{FR75} method, metal abundance is determined by plotting
the pseudo-equivalent width (EW) of the Ca II K line, $W$(K), against
the mean EW of the Balmer lines, $W$(H).  As the RRL varies in phase,
it traces out a path on this plot which is strongly dependent on its
metallicity.  Thus, by using a calibration based on RRLs of known
metallicity, we can determine [Fe/H] for our sample from low
resolution spectra.  Note that observations taken during rising light
(phase $\sim$ 0.8--1) should not be used since changes in the RRL's
effective gravity and Balmer line profiles during this stage alter the
relationship between $W$(K) and $W$(H).  We have also omitted type $c$
RRLs from the metallicity analysis since they are hotter and have
weaker Ca II K lines than type $ab$ RRLs.  Lower signal-to-noise
spectra and uncertainties in the contamination from interstellar Ca II
K would thus result in larger uncertainties in the metallicities of
type $c$ RRLs compared to type $ab$.

The first step was to normalize our wavelength-calibrated spectra to
unit intensity using IRAF's \textsc{continuum} task, which divides
each spectrum by an appropriate polynomial.  Subsequent steps closely
followed the method described in \citet{Layden94}.  Eight of
\citeauthor{Layden94}'s EW standard RRLs had been observed multiple
times over the course of our observing runs.  $W$(K) (corrected for
interstellar contamination using the \citealt{Beers90} model) and the
EWs of the Balmer lines H$\delta$, H$\gamma$ and H$\beta$ were
measured using numerical integration, with feature bands and continuum
bands equal to \citeauthor{Layden94}'s wherever possible (see Table
\ref{ewbands}).  Unlike \citeauthor{Layden94}, however, the $W$(K)
continuum bands were fixed for all reductions and our red continuum
band for H$\beta$ was truncated due to our smaller wavelength
coverage.  Also note that all bands were offset by an
appropriate wavelength shift according to the observed geocentric
radial velocity of each spectrum (described in \S\ref{rv}).

  The measured equivalent widths for the standard RRLs are shown in
  Table \ref{ewstds} and offsets from \citeauthor{Layden94}'s values
  are displayed in Figure \ref{resids}.  It can be seen that our
  values agree well with \citeauthor{Layden94}'s for $W$(K),
  $W$(H$\delta$) and $W$(H$\gamma$) but that our values for
  $W$(H$\beta$) are systematically smaller than
  \citeauthor{Layden94}'s.  This was likely due to the use of a
  different red continuum band as mentioned above.  A linear
  regression was performed to calculate the appropriate correction to
  bring our $W$(H$\beta$) measurements in line with
  \citeauthor{Layden94}'s.  The best fit is overplotted on Figure
  \ref{resids}.

%---------------------- resids --------------------------%
\begin{figure}[htbp]    
        \centering
        \epsscale{1.0}
        \plotone{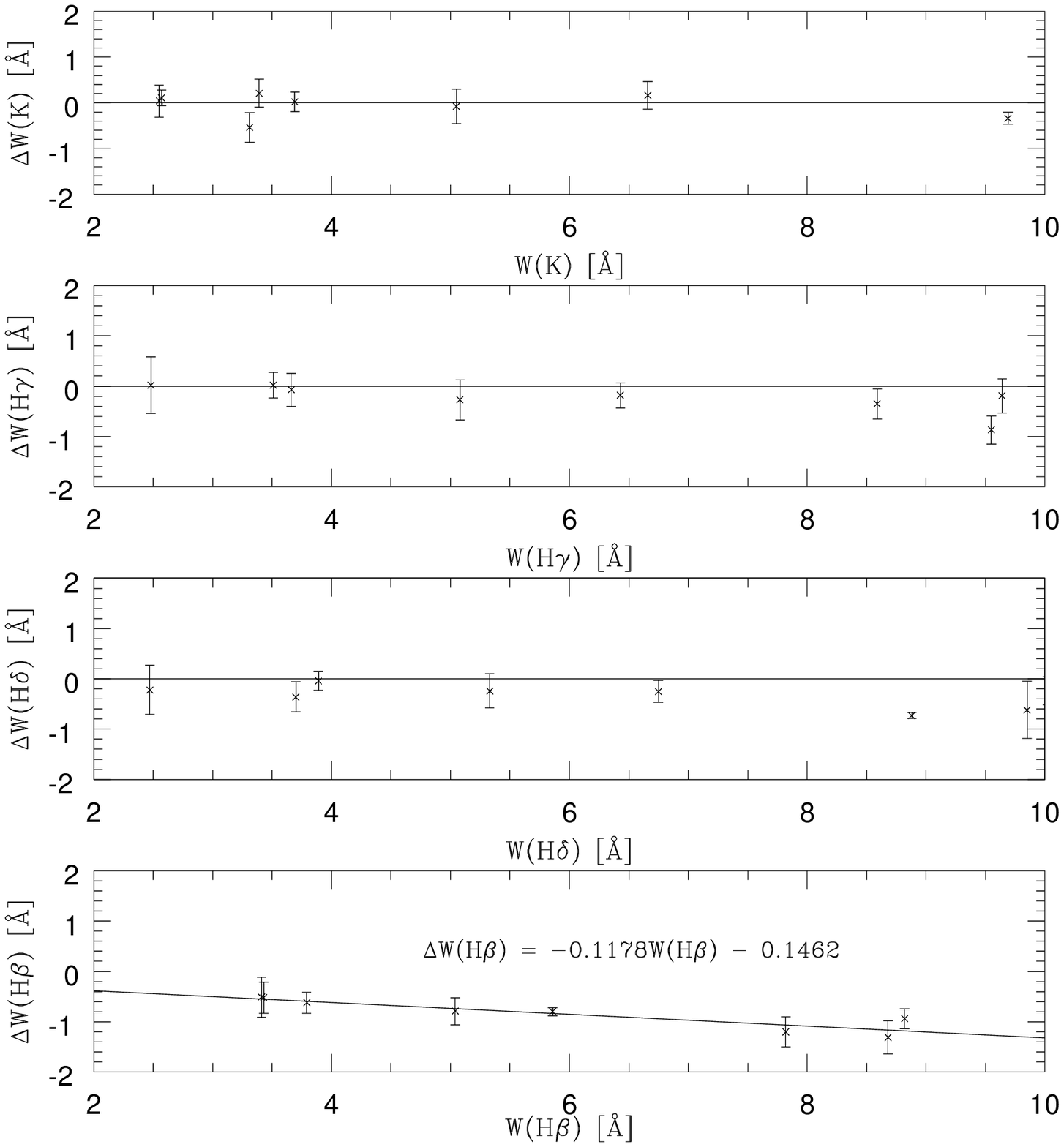}
        \caption[Differences between pseudo-equivalent widths in the
        present study and those of \citet{Layden94} for eight EW
        standard stars.]{Differences between pseudo-equivalent widths
        (\textit{top to bottom:} $W$(K), $W$(H$\gamma$),
        $W$(H$\delta$), $W$(H$\beta$)) in the current study and those
        of \citet{Layden94} for eight EW standard stars.  Negative
        $\Delta W$ values indicate that our EWs are smaller than those
        of Layden.  Error bars are the standard deviations of our
        repeated measures of the EWs (see Table \ref{ewstds} for
        details).  No appreciable difference was noted for $W$(K),
        $W$(H$\gamma$) or $W$(H$\delta$) (horizontal line drawn at
        $\Delta W$ = 0 for reference) but a negative linear
        relationship was noted for  $W$(H$\beta$), as shown by the
        solid line in the lowest plot.}  
        \label{resids}  
\end{figure}
%------------------------------------------------------%

%---------------------- h3k --------------------------%
\begin{figure}[htbp]    
        \centering
        \epsscale{0.8}
        \plotone{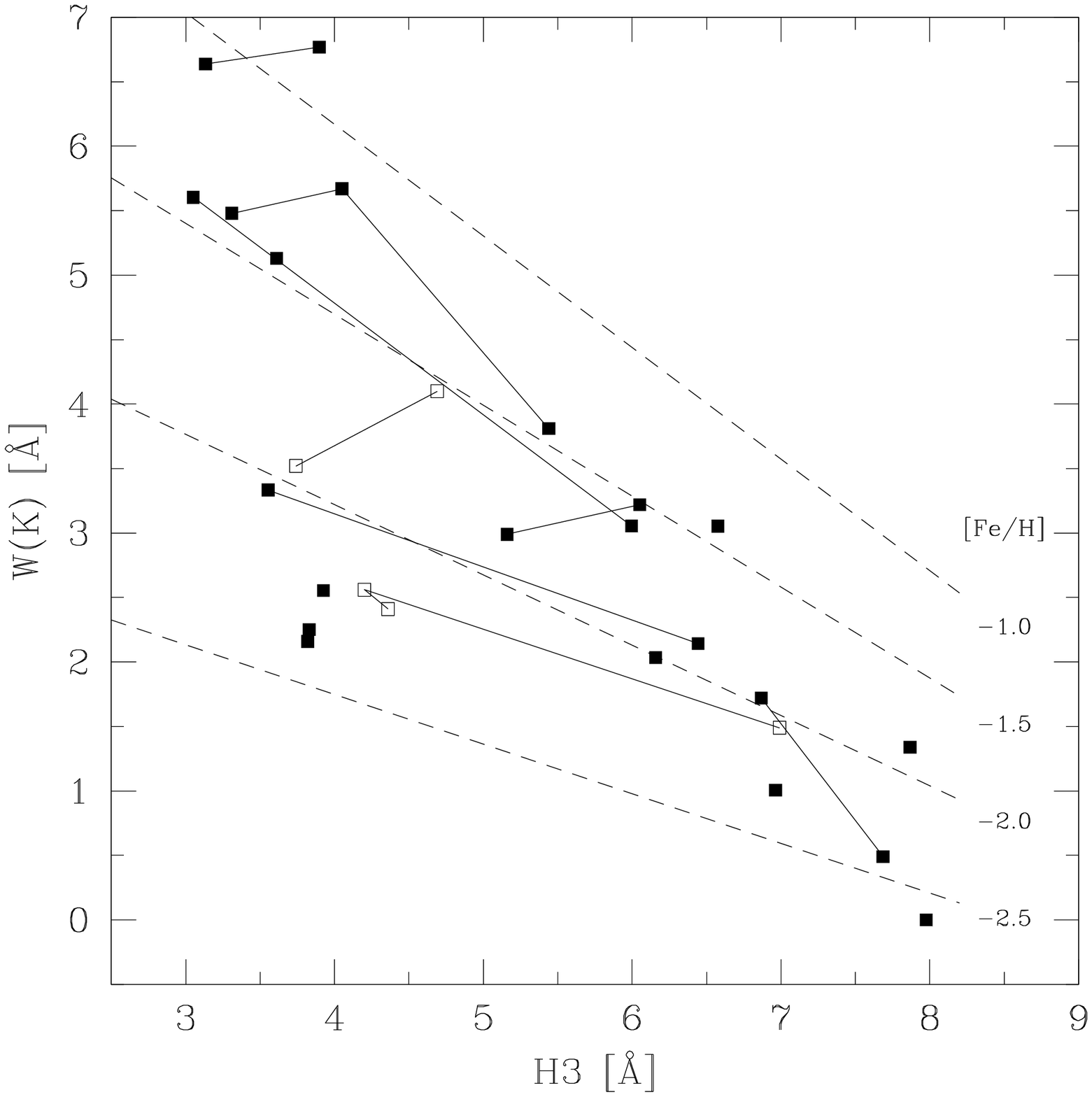}
        \caption[The pseudo-equivalent width of the Ca II K line,
        corrected for interstellar absorption, against the average
        width of H$\delta$, H$\gamma$ and H$\beta$ for the 16 type
        $ab$ RRLs in the 14 h, 16 h, 0 h, and VSS regions.]{The
        pseudo-equivalent width of the Ca II K line, corrected for
        interstellar absorption, against the average width of
        H$\delta$, H$\gamma$ and H$\beta$ for the 16 type $ab$ RRLs in
        the 14 h, 16 h, 0 h and VSS regions.  Solid lines connect
        values for the same RRL observed at different phases.  The
        dashed lines are the loci of stars having the indicated [Fe/H]
        value according to \citeauthor{Layden94}'s
        \citeyearpar{Layden94} calibration.  The two hypothesized VSS
        members are plotted as open squares.}  
        \label{h3k}     
\end{figure}
%------------------------------------------------------%

With the offsets to \citeauthor{Layden94}'s system in hand, the process
of normalization and measurement of EWs was repeated for the target
RRLs.  In Figure \ref{h3k}, $W$(K), corrected for interstellar Ca II
K, is plotted against $W$(H3), the average of the EWs of H$\delta$,
H$\gamma$ and H$\beta$ (offset to \citeauthor{Layden94}'s system).
The dashed lines are given by  

\begin{displaymath}
\textrm{$W$(K) = $a$ + $bW$(H) + $c$[Fe/H] + $dW$(H)[Fe/H],}
\end{displaymath}

\noindent using coefficients determined by \citet{Layden94} that yield
an external precision for [Fe/H] of $0.15$--$0.20$ dex.  The values of
[Fe/H] determined from this equation for the 16 type $ab$ RRLs are listed 
in Table \ref{spectro} and
their distribution is shown in Figure \ref{fehhisto}.  
Where more than one observation exists, the tabulated values were calculated 
by averaging the [Fe/H] values from the different phases (cf.\  Figure \ref{h3k}).
 Based on the stars with multiple observations, the 
internal precision of a single [Fe/H] determination is 0.20 dex.  For this sample 
\meanfeh $= -1.86 \pm 0.1$ with a dispersion $\sigma$ = 0.45
dex (see Figure \ref{fehhisto}).  This value is somewhat more metal
poor than that, \meanfeh $= -1.61 \pm 0.06$,
$\sigma$ = 0.4 dex, tabulated by \citet{KCC00} for RRLs in the halo.  The two
proposed VSS members (the other two members are type $c$ RRLs for
which metallicities could not be calculated) have \meanfeh $ = -1.95
\pm 0.1$ on our [Fe/H] system and an abundance range of $0.4 \pm 0.2$
dex.  These values agree with \meanfeh\sub{VSS} $= -1.86 \pm 0.08$,
$\sigma$ = 0.40 dex found by DZV06, supporting our claim that
these stars are part of the stream.  The two type $ab$ RRLs
in the negative \vgsr\ peak in Figure \ref{genhistovss}, which may be associated 
with Sgr debris, have \meanfeh $= -1.57 \pm
0.1$ and an abundance range of $0.2 \pm 0.1$ dex.

%---------------------- fehhisto --------------------------%
\begin{figure}[htbp]    
        \centering
        \epsscale{0.5}
        \plotone{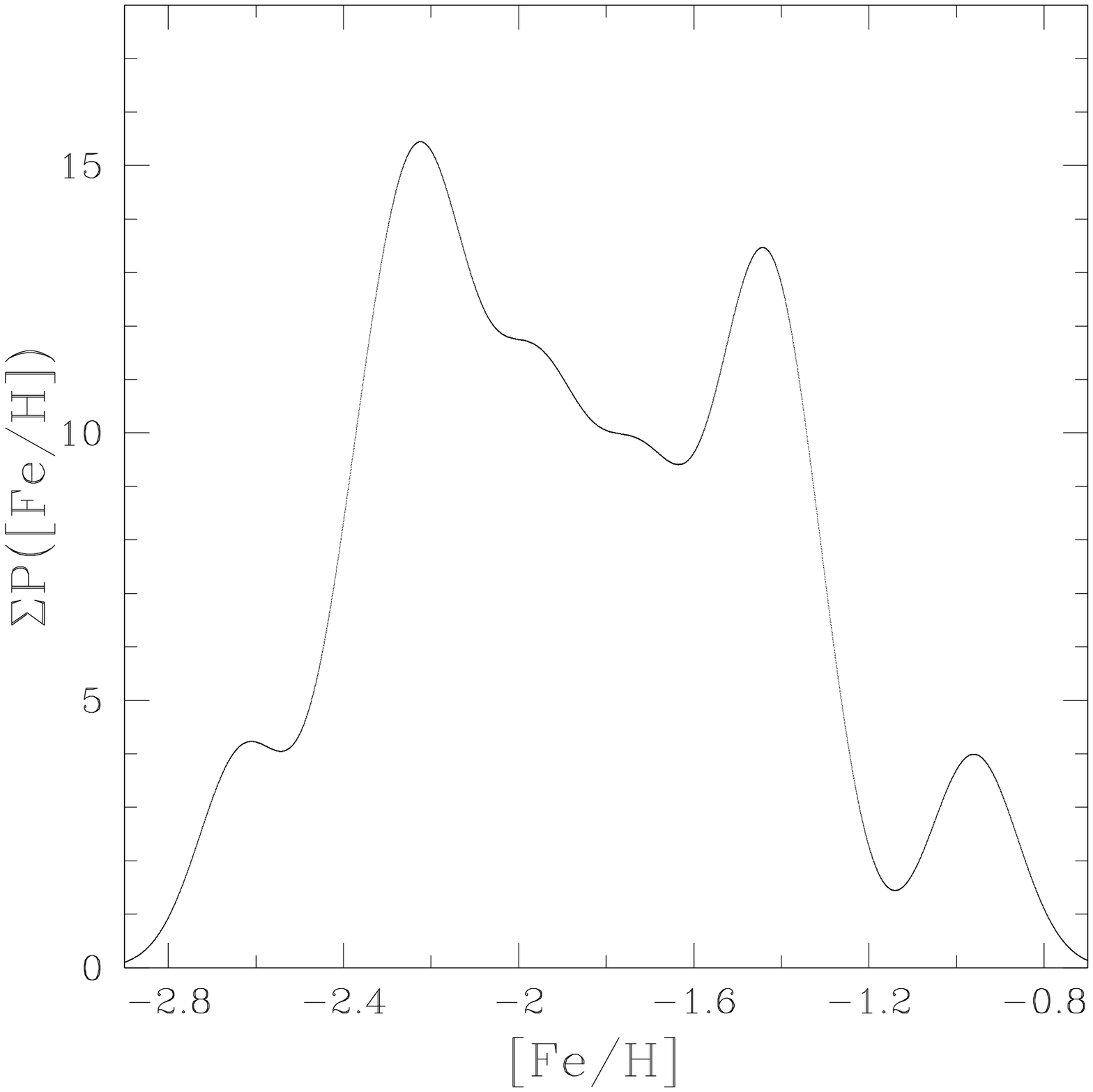}
        \caption{Generalized histogram of [Fe/H] (with kernel of 0.1
        dex) for the 16 type $ab$ RRLs in the 14 h, 16 h, 0 h, and VSS
        regions.}
        \label{fehhisto}        
\end{figure}
%------------------------------------------------------%

%\input{lf}

\section{Luminosity Functions}
\label{lf}

Having discovered three RRLs with radial velocities and metal
abundances consistent with the VSS yet falling outside the VSS region
identified by DZV06, it was of interest to further explore the spatial
extent of the stream by now examining a broader stellar  population.
We selected 49 2\degree$\times$2\degree{} regions spread over RA
125--220\degree{} with the aim of sampling the area roughly evenly,
given the constraints of the actual SEKBO field locations (see Figure
\ref{12hr4panel}).  Color-magnitude diagrams (CMDs) and luminosity
functions (LFs) were constructed from the SEKBO data (examples are
shown in Figure \ref{cmdlf}) and examined for signs of an upturn near
the magnitude where the subgiant branch and the main sequence merge.
For an old population, this occurs at $M_V \approx 3.5$ which corresponds
to $V \approx 19.9$ at a distance of 19 kpc (the average distance of the
four identified VSS members, consistent with the findings of
\citealt{NYR02} and DZV06).  This technique was also used by DZV06,
where the target region was compared to a control region of equal
area.  Given the difficulty in identifying a suitable control region
when the spatial extent of the VSS is unclear, we opted instead to
compare the observational data to synthetic data produced by the
Besan\c{c}on model of Galactic stellar populations \citep{RRD03}.
This model comprises four components: thin disk, thick disk, halo and
bulge.  It is a smooth, dynamically self-consistent model where
parameters are forced to follow physical laws, taking into account
physical links between density, velocity and metallicity distribution.

The simulations covered a distance interval of 0--120 kpc and assumed
an average interstellar extinction coefficient of $A_V$ = 0.75 mag
kpc\super{-1}.  This value was chosen so that the average integrated
line-of-sight extinction was in agreement with those derived using the
dust maps of \citet{SFD98} and it is close to the value suggested by
\citet{RRD03} for intermediate to high galactic latitudes (the region
studied here covers $30\degree \lesssim b \lesssim 60\degree$).
Initial cuts in magnitude ($13 < V < 22$) and color ($-0.5 < V-R <
1.5$) were made.  In order to omit local red dwarfs and focus more
clearly on the population of interest, only stars with $(V-R)_0 < 0.7$
were included in the luminosity functions.  Since the simulated color
interval greatly exceeded the observational color errors, the
simulations were not convolved with photometric errors.  Similarly,
completeness of the observations was not incorporated into the
simulated data since our analyses would focus on stars brighter than
the observational incompleteness limit.  

%-------------------- 12hr4panel --------------------------%
% BB 30 320 560 700
\begin{figure}[htbp]    
        \centering
        \epsscale{1.0}
        \plotone{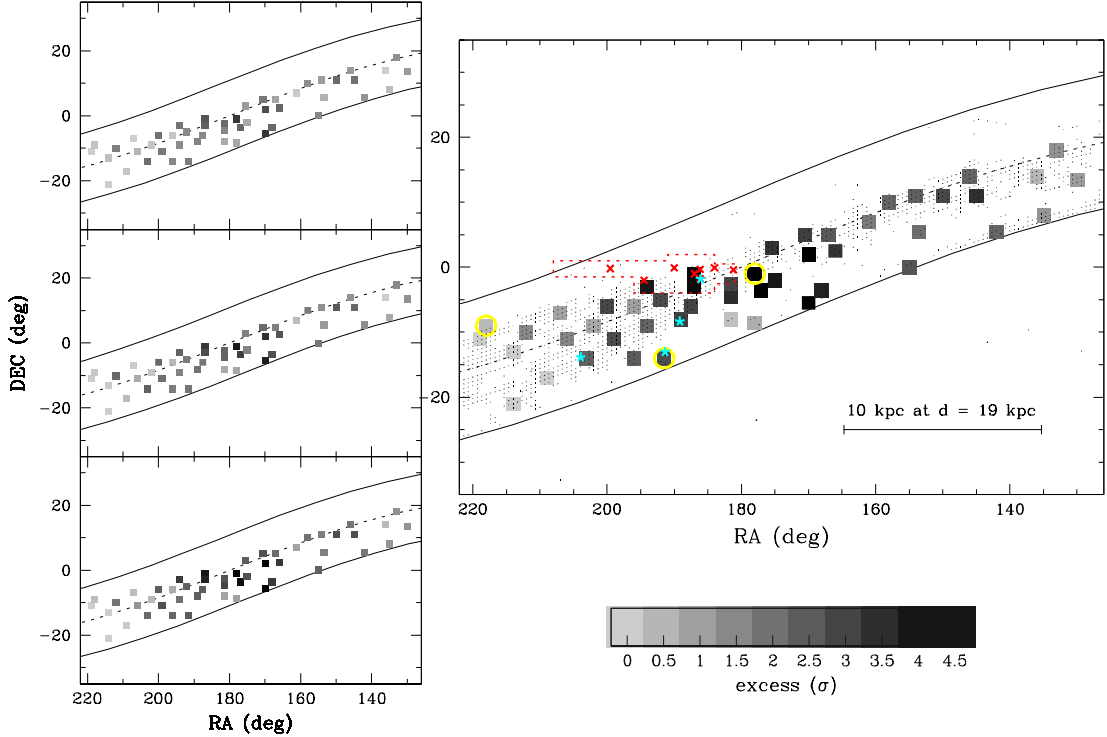}
        \caption[The spatial distribution of regions where data was
        compared to the model, with color representing the
        significance of the excess of data stars over synthetic
        stars.]{The spatial distribution of regions in which data was
        compared to the Galactic stellar distribution model.  The
        significance of the excess of data stars over synthetic stars
        is represented by the coloring of the region according to the
        scale shown, with darker shades indicating a more significant
        excess.  The range of magnitudes included in the excess
        calculation was $V_0$ = 16.25 to \textit{(left panel, top to
        bottom)} 18.25, 18.75, 19.25 or \textit{(right panel)} 19.75.
        Also shown are the four stars from our sample with \vgsr{}
        consistent with the VSS \textit{(cyan asterisks)} and those
        found by DZV06 in QUEST data \textit{(red crosses)}.  The
        example regions in Figure \ref{cmdlf} are circled in yellow.
        As in Figure \ref{12hrtno}, the red dashed box is DZV06's
        identification of the VSS region.  The dashed line represents
        the ecliptic and the solid lines outline the broad limits of
        the SEKBO survey, with small dots showing the centers of the
        fields surveyed.}
        \label{12hr4panel}      
\end{figure}
%------------------------------------------------------%

Equatorial coordinates of the regions simulated were identical to
those of the chosen observed regions (with a step size of 1\degree{}
in both RA and Dec), however, areas were not equal due to the
non-uniform sampling of the SEKBO survey (see the small dots in the
righthand panel of Figure \ref{12hr4panel}).  In order to compare the
LFs, we thus normalized the synthetic data to the observed data based
on counts in the range $14.7< V_0 < 16$.  As an example, Figure
\ref{cmdlf} shows CMDs and LFs for three regions.  In the upper panel,
a clear excess of observed stars over synthetic stars can be seen for
$V_0 > 16.5$, peaking at $V_0 \sim 19.5$.  It should be noted that
incompleteness becomes a significant factor by $V = 19$ and thus the
excess may well continue to grow to fainter magnitudes.  In this
particular region, two excesses are apparent in the CMD for stars
fainter than $V \sim 19$.  One has $(V-R)_0 \sim 0.35$ while the other
is redder, with $(V-R)_0 \sim 0.6$.  These excesses possibly
correspond to the top of the main sequence and the lower giant branch,
respectively.  We verified that the excess in the luminosity function
is still present using a bluer cutoff, $(V-R)_0 < 0.5$, and thus the
excess is not driven solely by the redder stars.  In the middle panel,
the overall excess is smaller, becoming noticeable only at $V_0 \sim
17.5$ and apparently peaking at $V_0 \sim 19$ before incompleteness
sets in.  There is thus some evidence for the VSS in the region
represented in the middle panel and evidence for a strong signal in
the region represented in the top panel.  The data in the bottom panel
follows the synthetic data closely, until dropping off at the faint
end when incompleteness sets in.  Such regions show no evidence of the
presence of the VSS.

As confirmation of these results, for those regions north of the declination limit of 
the SDSS we carried out a similar comparison using SDSS data rather 
than that from the SEKBO survey.
The SDSS data have the advantage of deeper limiting magnitude and complete area
coverage.  In Figure \ref{sdsslfcomp} we show comparisons similar to those
of Figure \ref{cmdlf}, but with the predictions of the Besan\c{c}on model now
compared to SDSS data, again for 2\degree$\times$2\degree{} regions.  
As for the SEKBO data, the color-magnitude diagrams and luminosity functions
in Figure \ref{sdsslfcomp} exclude local red dwarfs by considering only stars with 
$(g-r)_{0} < 1.0$, equivalent to $(V-R)_{0} \approx 0.7$.  The
upper panels of Figure \ref{sdsslfcomp} are in fact for the same region as the
upper panels of Figure \ref{cmdlf}, and, comfortingly, the results are very similar.
There is an increasing excess of stars above the model predictions with decreasing
magnitude that continues beyond the completeness limit of the SEKBO data,
confirming that the VSS is strongly present in this region.  The other two panels
show different regions to those in Figure \ref{cmdlf}, though they were selected in a similar 
way: the middle panel is for the region centered on (RA, Dec) = (170\degree, 
+2\degree) where the 
SEKBO data predicts the VSS is present (cf.\ Figure \ref{12hr4panel}) and the lower panel
is for the region (135\degree, +8\degree) where no significant excess is predicted.  In both instances the 
comparison with the SDSS data confirms the interpretation of the SEKBO data.

% strong VSS (top) -> field31
% weak VSS (middle) -> field42
% absent (bottom) -> field11

%-------------------- cmdlf --------------------------%
% BB 18 144 592 718
\begin{figure}[htbp]    
        \centering
        \epsscale{0.7}
        \plotone{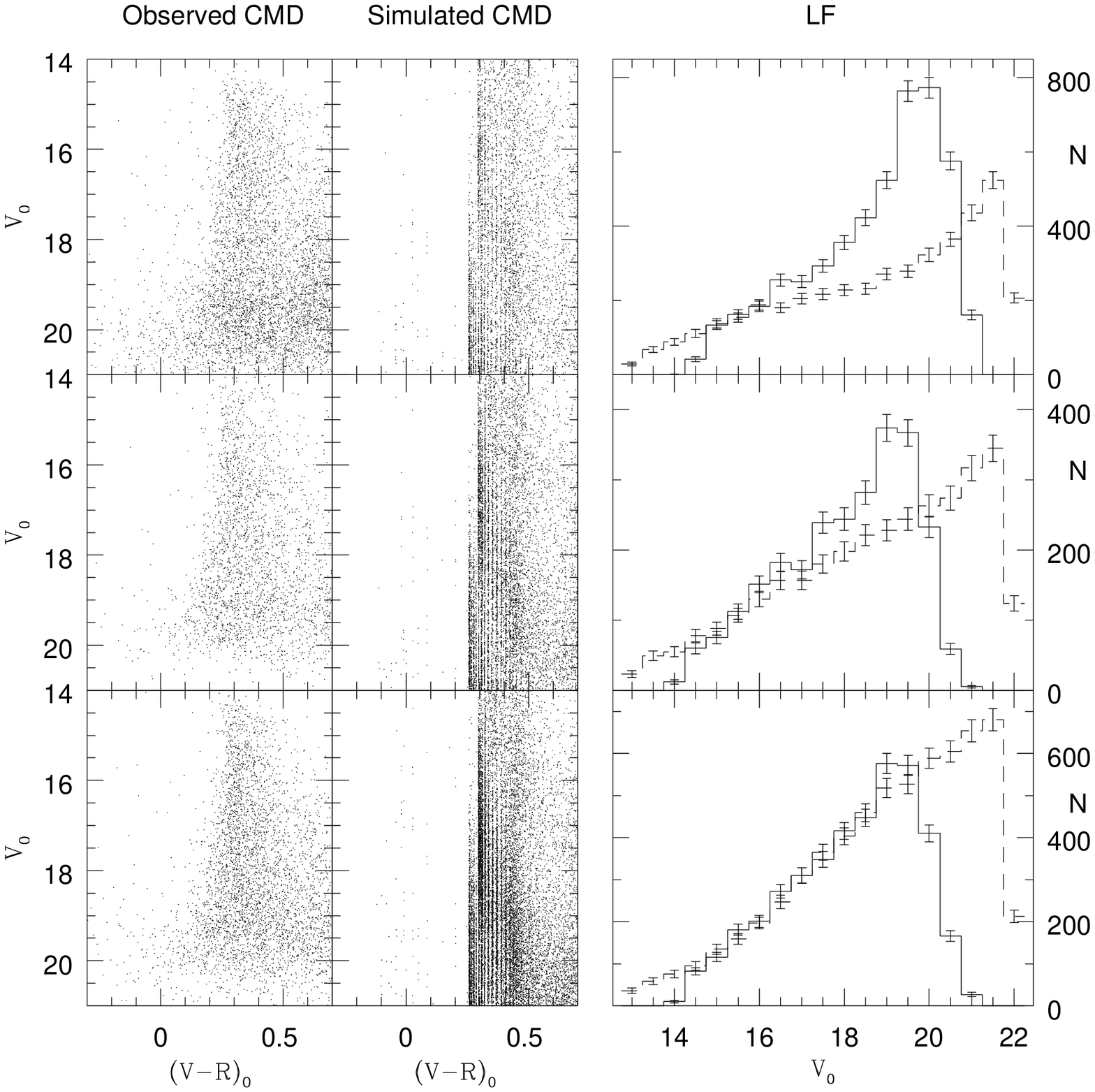}
        \caption[CMDs and LFs for three observed and synthetic
        fields.]{\textit{Left panels:} Example CMDs from observed data
        for 2\degree$\times$2\degree{} (non-uniformly sampled) regions
        (see yellow circles on Figure \ref{12hr4panel} for spatial
        locations).  \textit{Middle panels:} Corresponding CMDs from
        synthetic data using the Bescan\c{c}on model \citep{RRD03},
        not convolved with completeness or photometric errors.  Note
        that the areas have not been normalized in constructing the
        CMDs.  \textit{Right panels:} Luminosity functions for
        observed \textit{(solid line)} and synthetic data (normalized
        over $14.7<V_0<16$; \textit{dashed line}) in each region for
        stars with $(V-R)_0 < 0.7$.  Poisson error bars are shown.
        The field in the top panel (RA $\sim 178$\degree) shows a
        large excess of stars in the data compared to the model,
        providing strong evidence for the presence of the VSS.  The
        field in the middle panel (RA $\sim 192$\degree) shows a
        weaker excess, while the field in the bottom panel (RA $\sim
        218$\degree) shows no excess.}
        \label{cmdlf}   
\end{figure}
%------------------------------------------------------%

%-------------------------sdsslfcomp---------------%
\begin{figure}[htbp]
     \centering
     \epsscale{0.8}
   \plotone{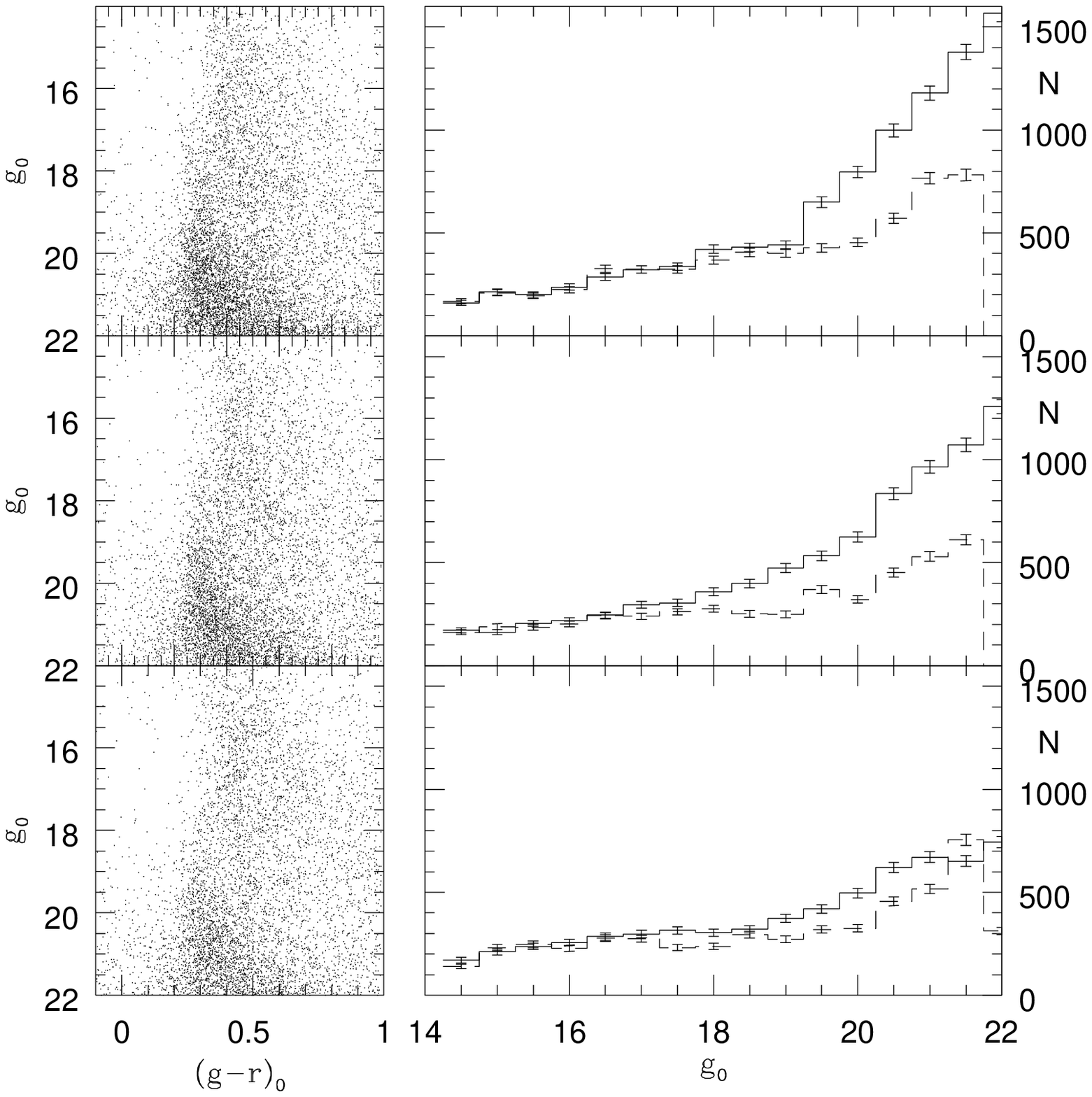}
    \caption[As for fig 13 but with SDSS data]{Color-magnitude diagram and luminosity
    function comparisons as for Figure \ref{cmdlf} but now with SDSS observational data
    rather than SEKBO data. The field for the upper panels is the same as in 
    Figure \ref{cmdlf}.  \textit{Left panels:} CMDs 
        for 2\degree$\times$2\degree{} regions from the SDSS.   
         \textit{Right panels:} Luminosity functions for
        observed \textit{(solid line)} and synthetic data (normalized
        over $14.9<g_0<16.2$; \textit{dashed line}) in each region for
        stars with $(g-r)_0 < 1.0$.  Poisson error bars are shown.
        As in Figure \ref{cmdlf} the top panel shows a
        large excess of stars compared to the model,
        providing strong evidence for the presence of the VSS.  The
        field in the middle panel (170\degree, +2\degree) also shows an
        excess, while the field in the bottom panel (135\degree, +8\degree) shows 
        reasonable agreement between the model and the observations.  These
        are consistent with the SEKBO data for these fields (cf.\ Figure \ref{12hr4panel}).}
        \label{sdsslfcomp}
 \end{figure}
 %-------------------------------------------------%

With this confirmation of the utility of the SEKBO data, we now examine the 
excess quantitatively, by computing for each
region the average difference between the observed data counts and the
normalized model counts between $V_0 = 16.25$ and $V_0 = $18.25,
18.75, 19.25 or 19.75.  Differences were calculated in 0.5 mag bins
and divided by the error in the difference, taken as the combined
Poisson error in the data and normalized model counts.  The average
error-weighted difference over the magnitude bins was then taken as a
measure of the significance of the excess in each region.  Figure
\ref{12hr4panel} displays the significance of the excess, represented
by the grayscale shading, as a function of spatial position.  The
regions colored black have a 4.5$\sigma$ excess of data stars over
model stars, providing strong evidence for the presence of the VSS in
those regions.  Since the excess appears to peak at $V_0\sim 19.5$ or
fainter, it is not surprising that using 19.75 as the faint limit of
the excess calculation (right panel) provides the highest sensitivity
to detection of the VSS.  The VSS signal is, however, present at
brighter magnitudes, albeit more weakly (left panels).  Taking into
account the scarcity of sampled fields to the west of RA $\sim
180\degree$, the overall pattern of excess significance is not
inconsistent with the location of overdensities (Clumps 1 and 2) of
the SEKBO survey RRL candidates at this distance.

While the foregoing analysis provided a significance map of the VSS,
it is also desirable to construct a density map so that the absolute
magnitude of the entire stream can be estimated.  To do this, we
scaled both observed and synthetic counts by the actual area covered.
After normalizing the synthetic data to the observed data based on
counts between $14.7 < V_0 < 16.0$ and subtracting synthetic data from
observed data, we then had a measure of the excess number of stars per
square degree in each 0.5 mag bin between $V_0 = 16.25$ and 19.75.
This excess number, summed over magnitude, is represented by the
grayscale shading in Figure \ref{12hrarea}, with black indicating a
500 star excess per square degree.  The overall pattern of excess is
similar to the significance map in the righthand panel of Figure
\ref{12hr4panel}, with perhaps the southern regions showing a stronger
signal in the density map than in the significance map.  This could be
understood in terms of the relative significance of the feature
decreasing towards the Galactic plane due to the increase in
background Milky Way stars, while the number density of stars in the
feature in fact remains constant.

%-------------------- 12hrarea --------------------------%
% BB 22 16 596 784  (rotate sm figure 3 times using epsffit)
\begin{figure}[htbp]    
        \centering
        \epsscale{1.0}  
        \plotone{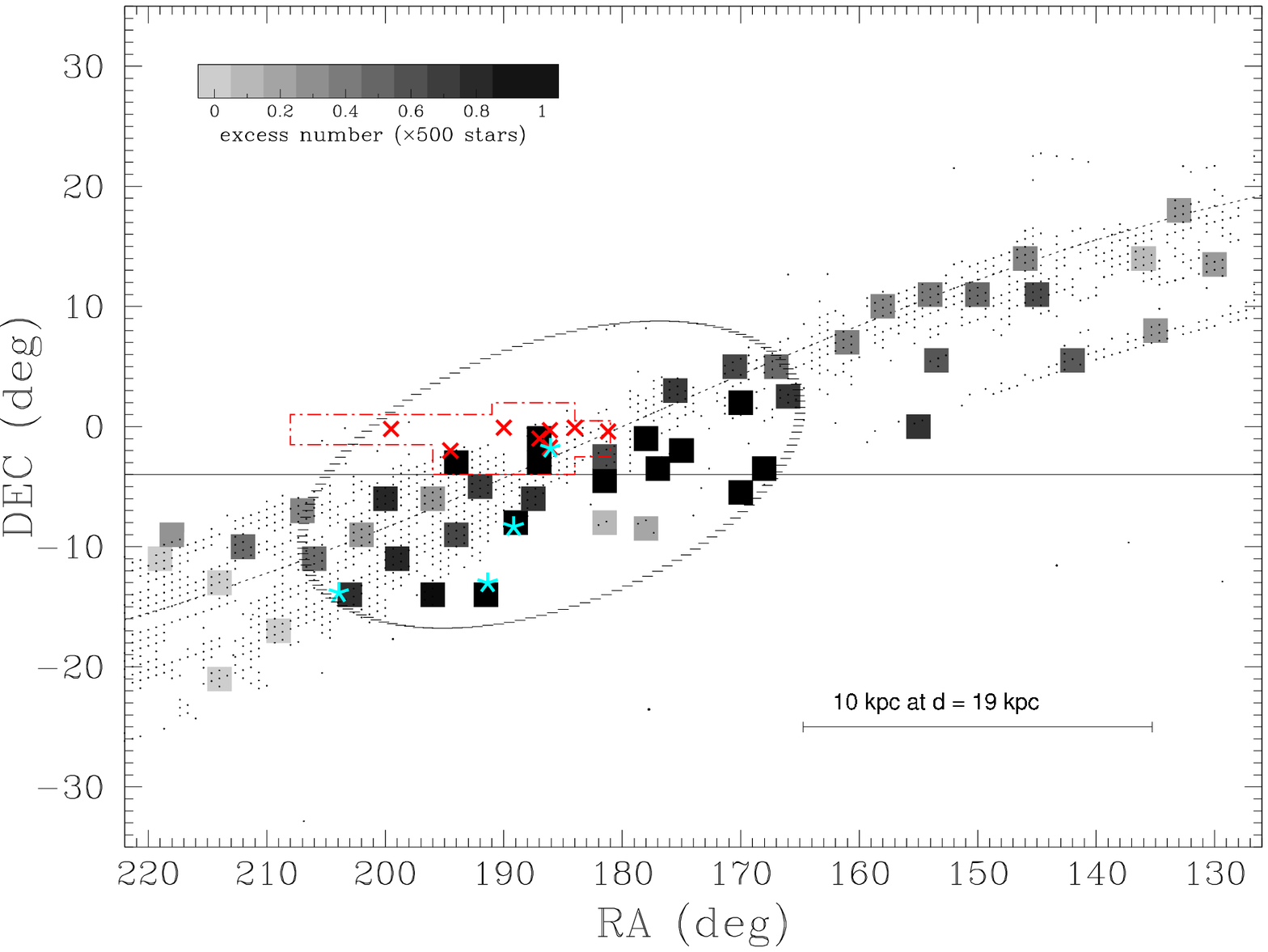}
        \caption[The spatial distribution of regions where data was
        compared to the model, with color representing the number of
        excess data stars over synthetic stars.]{The spatial
        distribution of regions in which data was compared to the
        Galactic stellar distribution model.  The number of excess
        data stars over synthetic stars is represented by the
        coloring of the region according to the scale shown, with
        darker shades indicating a larger excess.  The range of
        magnitudes included in the excess calculation was $16.25 < V_0
        < 19.75$.  Also shown are the four stars from our sample with
        \vgsr{} consistent with the VSS \textit{(cyan asterisks)} and
        those found by DZV06 \textit{(red crosses)}.  The ellipse
        outlines the estimated area of the VSS based on the current
        analysis.  As in Figure \ref{12hrtno}, the red dashed box is
        DZV06's identification of the VSS region.  The dashed line
        represents the ecliptic and the small dots show the centres of
        the fields surveyed.  The approximate southern limit of the
        SDSS coverage in this region is shown by the solid line at Dec =
        $-4$\degree.}  
        \label{12hrarea}        
\end{figure}
%------------------------------------------------------%

In order to make a rough estimation of the sky coverage of the VSS, an
ellipse has been overplotted on Figure \ref{12hrarea}.  The chosen
shape is somewhat arbitrary, with the positioning and size selected so
as to include the regions in which the excess appears visually to be
significant.  The ellipse encompasses areas not sampled by the SEKBO
survey on the basis that the VSS could plausibly extend to those areas
given the distribution of high excess regions in the sampled areas and
assuming a certain degree of uniformity.  This assumption could lead
to an overestimate in the area, but conversely, the VSS may well
extend beyond the survey region (particularly to the south where no
data has been gathered by previous surveys) leading to an area
underestimate.  Entire coverage will be possible with SkyMapper
\citep{KSB07}, but in the meantime we note that the current area
estimate is likely to be uncertain by at least a factor of two.  

Assuming a heliocentric distance of 19 kpc, the stream's projection on
the sky extends $\sim$15 kpc in length along the largest dimension and
covers an area of $\sim$760 deg\super{2}, centered at (RA, Dec) $\sim
(186\degree, -4\degree)$.  This estimate of the area, with its large
uncertainty, is in rough agreement with \citeauthor{JIB08}'s
\citeyearpar{JIB08} estimate of 1000 deg\super{2} for the VSS based on
SDSS data.  Not only is there evidence that the feature is much larger
than DZV06's estimate of 106 deg\super{2} (dashed box on Figure
\ref{12hr4panel}), but given the non-overlapping regions of the SDSS
and SEKBO survey data, the true size of the stream may well be greater
than 1000 deg\super{2}.  Our analysis shows the VSS to extend to the
west and to the south of DZV06's detection.  We indeed found three
RRLs in the southern direction with radial velocities consistent with
the VSS.  Note that neither QUEST nor SDSS covered Galactic latitudes
lower than $\sim$60\degree{} in this region, hence our findings
provide the first tracing of this section of the stream, extending the
location of the VSS well south of the declination limits of these two surveys.
\citet{JIB08} found that the number density of stars belonging to the
feature increases towards the Galactic plane and indeed we find that
the VSS extends to at least $b\sim45\degree$ (Dec $\sim -15$\degree).

Having estimated the area of the VSS, it is now possible to calculate
an estimate of the absolute magnitude, $M_V$, of the stream.  The
fluxes for the excess stars in each magnitude bin ($16.25<V_0<19.75$)
were averaged over all the regions contained within the ellipse,
summed over the magnitude bins and finally, multiplied by the area of
the ellipse.  Assuming a distance of 19 kpc and an area of 760
deg\super{2} and using $V$ band values of $M_{\odot} = 4.83$ and
$L_{\odot} = 4.64\times10^{25}$ W \citep{BM98}, we calculate $M_V =
-11.9$ mag.  This value is considerably brighter than $M_R = -8.0$ mag
estimated by \citet{JIB08}.  We note, however, that their value
assumed a distance of 10 kpc, an area of 1000 deg\super{2} and
magnitude limits of $18 < R < 21.5$.  Using values as close to these
as possible given the constraints of our data, our estimate becomes
$M_V = -10.1$ mag.

A final point to note is that \citet{JIB08} find the VSS to span
$\sim$10 kpc along the line-of-sight, at distances closer than where
it was detected by the SEKBO survey, and that their survey data did
not go beyond a scale height of $Z = 20$ kpc.  It thus seems a likely
scenario that the VSS spans many kpc along the line-of-sight (indeed
DZV06 found possible members at distances ranging between 16 and 24
kpc and new results of \citealt{VJZ08} find possible members as close
as 12 kpc) but is more diffuse at distances $\lesssim$15 kpc, with the
highest concentration at $d \sim 19$ kpc.  Considering that
\citet{JIB08} do not include the portion at $d \sim 19$ kpc in their
$M_R$ calculation, it is not surprising that their value should be
considerably fainter than ours.  It is also important to note that all
our values are lower limits only, since stars brighter than $V_0 =
16.25$ and fainter than $V_0 = 19.75$ were not included.  In addition,
incompleteness was not taken into account.  The estimate is also
sensitive to the area covered, distribution of VSS density within that
area and to the distance of the stream, each of which are somewhat
uncertain based on the sampling of the data currently available and
the likely extended nature of the stream along the line-of-sight.

Nevertheless, the VSS is clearly a significant local structure.  Its
origin remains unclear, though the large abundance range observed
($\sim$0.4 dex) 
is consistent with DZV06's suggestion that the VSS is the disrupted
remnants of a dwarf Spheroidal galaxy.  It is certainly a large,
diffuse structure and is likely to have a substantial total
luminosity.  Future kinematic observations are needed to further
constrain the properties of the system and to provide additional clues
to its origin.

\section{Conclusions}
\label{conc}

Analysis of follow-up spectroscopy of eleven photometrically confirmed
RRLs from a candidate list based on SEKBO survey data has revealed
three new RRLs with velocities consistent with membership in the Virgo
Stellar Stream, in addition to one previously identified member
(\meanvgsr = $127 \pm 10$ \kms{}, $\sigma = 27$ \kms).  The two type
$ab$ members have \meanfeh{} = $-1.95 \pm 0.1$ and an abundance range
of $\sim$0.4 dex, consistent with values found by DZV06 for the VSS\@.
The newly discovered VSS members occupy a region of space covered by
neither QUEST nor SDSS data, to the south-east of the apparent center
of the stream at (RA, Dec) $\sim$ ($186\degree,-4\degree$).
Comparison of luminosity functions for observed data compared to data
synthesized with the Besan\c{c}on Galactic model \citep{RRD03}
revealed the VSS to be a large, diffuse feature, covering at least 760
deg$^2$ of sky. The core of the VSS appears to have an angular size of
$\sim$45\degree{} along the longest dimension, corresponding to a
spatial scale of $\sim$15 kpc in projection, assuming a heliocentric
distance of $\sim$19 kpc.  We have traced the stream as far south as
Dec $\sim -14$\degree{} and to Galactic latitudes as low as $b \sim
45$\degree.

\acknowledgments

We are grateful to Mike Bessell for generously taking time out of his
own 2.3m run to take spectroscopic observations of two suspected VSS
members which would otherwise have been lost due to poor weather.
This research has been supported in part by the Australian Research
Council through Discovery Project Grants DP0343962 and DP0878137.

%\url{http://www.aas.org/publications/aastex}.
%\email{aastex-help@aas.org}.

%% See the AASTeX Web site at http://www.journals.uchicago.edu/AAS/AASTeX
%% for information on obtaining the facility keywords.
%{\it Facilities:} \facility{Nickel}, \facility{HST (STIS)}, \facility{CXO (ASIS)}.

{\it Facilities:} \facility{SSO:1m (WFI)}, \facility{SSO:2.3m (DBS-B)}

%---------------------- targets --------------------------%
\begin{deluxetable}{lc}
\tablewidth{0pt}
\tabletypesize{\small}
\tablecaption{Target Summary\label{targets}}
\tablehead{
\colhead{}       & \colhead{Number of Targets}}
\tablecolumns{2}
\startdata
\multicolumn{2}{c}{\textsc{Photometry and Spectroscopy}}   \\ 
\textbf{VSS:}                    &                         \\
\phm{abcdefg}Clump 1 (12.4 h)    &                       8 \\
\phm{abcdefg}Clump 2 (14 h)      &                       3 \\
\textbf{Sgr:}                    &                         \\
\phm{abcdefg}Clump 1 (20 h)      &                       5 \\
\phm{abcdefg}Clump 2 (21.5 h)    &                      21 \\   
\textbf{Other clumps:}           &                         \\           
\phm{abcdefg}14 h                &                       3 \\
\phm{abcdefg}16 h                &                       6 \\
\phm{abcdefg}0 h                 &                       5 \\
\textit{Spectroscopy Total}      &              \textit{51}\\
\\
\multicolumn{2}{c}{\textsc{Photometry Only}}               \\
\textbf{Contamination check:}    &                         \\
\phm{abcdefg}0 -- 21.5 h          &                     55 \\
\textit{Photometry Total}        &             \textit{106}\\   
\enddata
\end{deluxetable}
%---------------------------------------------------------%

%---------------------- ewbands --------------------------%
\begin{deluxetable}{lccccccc}
\tablewidth{0pt}
\tabletypesize{\small}
\tablecaption{Feature and continuum band wavelengths (in \AA)\label{ewbands}}
\tablehead{
\colhead{Feature}&\colhead{$\lambda$\sub{feature}}
&\multicolumn{2}{c}{Feature Band}
&\multicolumn{2}{c}{Blue Contin. Band}
&\multicolumn{2}{c}{Red Contin. Band}
\\
\multicolumn{2}{l}{}
&\colhead{$\lambda$\sub{blue}} &\colhead{$\lambda$\sub{red}}
&\colhead{$\lambda$\sub{blue}} &\colhead{$\lambda$\sub{red}}
&\colhead{$\lambda$\sub{blue}} &\colhead{$\lambda$\sub{red}}
}
\tablecolumns{8}
\startdata
Ca II K-narrow & 3933.666& 3927.& 3941.& 3908.& 3923.& 4019.& 4031.\\
Ca II K-wide   & 3933.666& 3924.& 3944.& 3908.& 3923.& 4019.& 4031.\\
H$\delta$      & 4101.735& 4092.& 4112.& 4008.& 4060.& 4140.& 4215.\\
H$\gamma$      & 4340.465& 4330.& 4350.& 4206.& 4269.& 4403.& 4476.\\
H$\beta$       & 4861.327& 4851.& 4871.& 4719.& 4799.& 4925.& 4950.\\
\enddata
\end{deluxetable}
%---------------------------------------------------------%

%---------------------- ewstds --------------------------%
\begin{deluxetable}{lccccccccc}
\tablewidth{0pt}
\tabletypesize{\small}
\tablecaption{Standard equivalent widths (in \AA)\label{ewstds}}
\tablehead{
\colhead{Star}&\colhead{N\sub{obs}}
&\multicolumn{2}{c}{$W$(K)}
&\multicolumn{2}{c}{$W$(H$\delta$)}
&\multicolumn{2}{c}{$W$(H$\gamma$)}
&\multicolumn{2}{c}{$W$(H$\beta$)}
\\
\multicolumn{2}{l}{}
&\colhead{mean} &\colhead{sd}
&\colhead{mean} &\colhead{sd}
&\colhead{mean} &\colhead{sd}
&\colhead{mean} &\colhead{sd}
}
\tablecolumns{10}
\startdata
BD-17 484 &  8& 5.05\super{w}& 0.09&  3.70& 0.05& 3.51& 0.07& 3.41& 0.07\\
HD 22413  & 10& 3.69\super{n}& 0.05&  6.75& 0.05& 6.43& 0.05& 5.86& 0.06\\
HD 65925  &  2& 6.66\super{w}& 0.04&  5.33& 0.06& 5.08& 0.07& 5.04& 0.01\\
HD 74000  &  4& 3.31\super{n}& 0.10&  3.89& 0.02& 3.66& 0.06& 3.43& 0.17\\
HD 74438  &  2& 3.39\super{n}& 0.00&  8.88& 0.01& 8.59& 0.08& 7.82& 0.01\\
HD 76483  &  7& 2.55\super{n}& 0.11&  9.85& 0.13& 9.55& 0.17& 8.68& 0.22\\
HD 78791  &  2& 9.69\super{w}& 0.03&  2.47& 0.07& 2.48& 0.06& 3.79& 0.07\\
HD 180482 &  4& 2.57\super{n}& 0.02& 10.02& 0.07& 9.64& 0.07& 8.82& 0.10\\
\tableline
\multicolumn{2}{l}{mean standard deviation}& & 0.06& & 0.06& & 0.08& & 0.09\\
\enddata
\tablecomments{When $W$(K\protect{\sub{wide}}) $< 4.0$ \AA, the narrow
  (n) feature band was used; when $W$(K\protect{\sub{wide}})
  \protect{$\geqslant$} 4.0 \AA, the wide (w) feature band was used.} 
\end{deluxetable}
%---------------------------------------------------------%

%\appendix

%\input{photdatashort}

\begin{deluxetable}{lcrccccccc}
\rotate
\tabletypesize{\small}
\tablewidth{0pt}
\tablecaption{Photometric data summary\label{photdata}}
\tablehead{

\colhead{ID} & \colhead{$\alpha$ (J2000.0)} & \colhead{$\delta$ (J2000.0)}  & \colhead{$\langle$$V$$\rangle$}  & \colhead{$\langle$$R$$\rangle$} & \colhead{$\langle$$V-R$$\rangle$$_0$} & \colhead{n\sub{obs}} & \colhead{Classification} & \colhead{Period (days)} & \colhead{$V$ Amplitude}}

\startdata

96102-170 & 12 24 15.67 & $-$01 49 14.52 & 17.02 & 16.83 & 0.174 & 15 & RR$ab$ & 0.525 & 0.286 \\
96637-458 & 08 48 02.53 & 17 02 45.92 & 17.19 & 16.68 & 0.499 & 6 & non-variable & - & - \\
97883-402 & 13 51 34.58 & $-$11 46 40.63 & 17.30 & 17.06 & 0.199 & 5 & RR$ab$ & 0.542 & 0.718 \\
97890-199 & 13 43 04.44 & $-$11 34 05.74 & 16.34 & 16.18 & 0.130 & 6 & RR$c$ & 0.350 & 0.374 \\
97890-1542 & 13 45 14.61 & $-$11 12 09.57  & 17.24 & 17.09 & 0.105 & 7 & unclassified variable & - & - \\
99747-73 & 21 14 09.39 & $-$17 36 32.90 & 17.09 & 16.91 & 0.135 & 8 & RR$ab$ & 0.517 & 0.724 \\
99752-96 & 21 33 35.19 & $-$16 07 05.52 & 17.08 & 16.93 & 0.098 & 15 & RR$c$ & 0.324 & 0.466 \\

\enddata

\tablecomments{Table \ref{photdata} is published in its entirety in the electronic version of the Journal.  A portion is shown here for guidance regarding its form and content.  Units of right ascension are hours, minutes, and seconds, and units of declination are degrees, arcminutes, and arcseconds.}

\end{deluxetable}

\begin{deluxetable}{lcccrrcc}
\tablewidth{0pt}
\tabletypesize{\small}
\tablecaption{Spectroscopic data summary\label{spectro}}
\tablehead{
\colhead{ID}   &\colhead{RRL}   &\colhead{systemic vel.}
&\colhead{n\sub{vel}}   &\colhead{$V$\sub{helio}}   &\colhead{\vgsr}
&\colhead{n\sub{Fe/H}}   &\colhead{[Fe/H]}
\\
&\colhead{type}   &\colhead{calculation}
&\colhead{}   &\colhead{(\kms)}   &\colhead{(\kms)}
&\colhead{}   &\colhead{}
}

\startdata

\multicolumn{8}{l}{\textit{VSS region}}\\
96102-170*  & $ab$ & fit     & 2 & $230$ &  $128$  &  3 &  $-2.15$ \\
105648-222  & $ab$ & fit     & 1 & $-91$ &  $-162$ &  1 &  $-1.45$ \\
107552-323  & $ab$ & fit     & 3 & $302$ &  $193$  &  3 &  $-1.34$ \\
108227-529  & $c$  & average & 1 & $-8$  &  $-119$ &  - &     - \\
109247-528  & $c$  & average & 2 & $113$ &  $28$   &  - &     - \\
119827-670  & $ab$ & fit     & 2 & $-84$ &  $-192$ &  2 &  $-1.68$ \\
120185-77   & $ab$ & fit     & 2 & $94$  &  $ 1$   &  2 &  $-2.38$ \\
120679-336* & $ab$ & fit     & 3 & $204$ &  $91$   &  2 &  $-1.74$ \\
120698-392* & $c$  & average & 1 & $227$ &  $134$  &  - &     - \\
121242-188* & $c$  & average & 3 & $276$ &  $155$  &  - &     - \\
121194-205  & $c$  & average & 1 & $-39$ &  $-152$ &  - &     - \\
\\
\multicolumn{8}{l}{\textit{14 h, 16 h, and 0 h regions}}\\
97890-199   & $c$  & average & 2 &     $367$ &    $285$ & - & -\\
106586-211  & $ab$ & fit     & 2 &      $-5$ &    $-33$ & 2 & $-1.52$\\
114421-242  & $ab$ & fit     & 1 &     $-51$ &     $40$ & 1 & $-2.30$\\
120857-475  & $ab$ & fit     & 1 &     $191$ &    $100$ & 1 & $-2.24$\\
121817-2385 & $ab$ & fit     & 2 &       $3$ &    $-33$ & 1 & $-1.87$\\
121906-2336 & $c$  & average & 2 &     $40$  &    $-2$  & - & -\\
122112-595  & $ab$ & fit     & 1 &    $-134$ &   $-168$ & 1 & $-1.42$\\
122156-1114 & $c$  & average & 2 &     $-42$ &    $-81$ & - & -\\
122214-1315 & $ab$ & fit     & 1 &      $24$ &    $-14$ & 2 & $-1.96$\\
122240-33   & $ab$ & average & 2 &     $121$ &     $21$ & 1 & $-2.01$\\
127008-210  & $ab$ & fit     & 2 &      $-3$ &     $78$ & 2 & $-0.96$\\
127806-85   & $ab$ & fit     & 1 &    $-187$ &   $-102$ & 1 & $-2.63$\\
127806-438  & $c$  & average & 3 &    $-165$ &    $-80$ & - & - \\
128416-544  & $ab$ & fit     & 2 &    $-309$ &   $-192$ & 2 & $-2.19$\\

\enddata

\tablecomments{Proposed VSS members are marked by *}

\end{deluxetable}

%% \bibliographystyle{apj}
%% \bibliography{apj-jour,ms}

\end{document}